\documentclass[twocolumn,aps,showpacs,amssymb,floatfix]{revtex4}
\newcommand{\be}{\begin{equation}}
\newcommand{\ee}{\end{equation}}
\newcommand{\beq}{\begin{eqnarray}}
\newcommand{\eeq}{\end{eqnarray}}

\usepackage{bm}
\usepackage{graphicx}%
\usepackage{epsf}
\usepackage{epsfig}

 \usepackage{amsmath}
\usepackage{amsfonts,amssymb}

\usepackage{dsfont}
\usepackage{slashed}
\usepackage{booktabs}

\begin{document}

\title{Nucleon form factors and moments of generalized  parton distributions
using $N_f=2+1+1$ twisted mass fermions}


\author{ C.~Alexandrou~$^{(a,b)}$,
  M. Constantinou~$^{(a)}$, S. Dinter~$^{(c)}$, V. Drach~$^{(c)}$,
  K.~Jansen~$^{(a,c)}$, C. Kallidonis~$^{(a)}$, G. Koutsou$^{(b)}$}
\affiliation{$^{(a)}$ Department of Physics, University of Cyprus, P.O. Box 20537, 1678 Nicosia, Cyprus\\
 $^{(b)}$ Computation-based Science and Technology Research
    Center, Cyprus Institute, 20 Kavafi Str., 2121 Nicosia, Cyprus \\
$^{(c)}$ NIC, DESY, Platanenallee 6, D-15738 Zeuthen, Germany}
         \vspace{0.2cm}

\begin{abstract}

We present results on the axial and the electromagnetic form factors of the nucleon, as well as, on
the first moments of the nucleon generalized parton distributions
using  maximally twisted mass fermions. We analyze
two $N_f{=}2{+}1{+}1$ ensembles having pion masses of 
213~MeV  and 373~MeV each at a different value of the lattice spacing. The lattice scale is
determined using the nucleon mass computed
 on a total of 17  $N_f{=}2{+}1{+}1$  ensembles generated at three
values of the lattice spacing, $a$. The renormalization constants are
evaluated non-perturbatively with a perturbative subtraction
of ${\cal O}(a^2)$-terms.
The moments of the generalized parton distributions are
given in the $\overline{\rm MS}$ scheme at a scale of $ \mu=2$~GeV.
We compare  with recent results obtained using different
discretization schemes. The implications on  the spin content of the
nucleon are also discussed.
\end{abstract}


\maketitle

\setcounter{figure}{\arabic{figure}}

\newcommand{\Op}{\mathcal{O}} 
\newcommand{\PO}{\mathcal{P}} 
\newcommand{\C}{\mathcal{C}} 
\newcommand{\eins}{\mathds{1}} 
\newcommand{\J}{\mathcal{J}}
\newcommand{\Dlr}{\buildrel \leftrightarrow \over D\raise-1pt\hbox{}}

\newcommand{\twopt}[5]{\langle G_{#1}^{#2}(#3;\mathbf{#4};\Gamma_{#5})\rangle}
\newcommand{\threept}[7]{\langle G_{#1}^{#2}(#3,#4;\mathbf{#5},\mathbf{#6};\Gamm
a_{#7})\rangle}

\bibliographystyle{apsrev}                     

\section{Introduction}

Recent progress in the numerical simulation of Lattice Quantum
Chromodynamics (LQCD) has been remarkable. The improvements in the
algorithms used and the increase in computational power have enabled
simulations to be carried out at near physical parameters of the
theory. This opens up exciting possibilities for {\it ab initio}
calculation of experimentally measured quantities, as well as, for
predicting quantities that are not easily accessible to
experiment. Understanding nucleon structure from first principles is
considered a milestone of hadronic physics and a rich experimental
program has been devoted to its study, starting with the measurements
of the electromagnetic form factors initiated more than 50
years ago. Reproducing these key observables within the LQCD
formulation is a prerequisite to obtaining reliable predictions on
observables that explore Physics beyond the standard model.

A number of major collaborations have been studying nucleon structure
within LQCD for many years. However, it is only recently that these
quantities can be obtained with near physical parameters both in
terms of the value of the pion mass, as well as, with respect to the
continuum limit~\cite{Alexandrou:2011nr, Alexandrou:2011db,
  Alexandrou:2010hf, Syritsyn:2009mx, Yamazaki:2009zq, Bali:2012av,
  Green:2012ud, Sternbeck:2012rw, Collins:2011mk, Brommel:2007sb,
  Hagler:2007xi}. The nucleon electromagnetic form factors are a well
suited experimental probe for studying nucleon structure and thus
provide a valuable benchmark for LQCD. The nucleon form factors
connected to the axial-vector current are more difficult to measure
and therefore less accurately known than its electromagnetic form
factors. A notable exception is the nucleon axial charge, $g_A$, which is
accurately measured in $\beta$-decays. The fact that $g_A$ can be extracted
at zero momentum transfer and that it is  technically straight forward to
be computed in LQCD, due to its isovector nature, makes it an ideal
benchmark quantity for LQCD. The Generalized Parton
Distributions (GPDs) encode information related to nucleon structure
that complements the information extracted from form
factors~\cite{Mueller:1998fv,Ji:1996nm,Radyushkin:1997ki}. They enter
in several physical processes such as Deeply Virtual Compton
Scattering and Deeply Virtual Meson Production. Their forward limit
coincides with the usual parton distributions and, using Ji's sum
rule~\cite{Ji:1996ek}, allows one to determine the contribution of a
specific parton to the nucleon spin. In the context of the
``proton spin puzzle'', which refers to the unexpectedly small fraction of
the total spin of the nucleon carried by quarks, this has triggered
intense experimental activity~\cite{Airapetian:2009rj, Chekanov:2008vy, Aaron:2007cz,
MunozCamacho:2006hx, Stepanyan:2001sm}.

\section{Lattice evaluation}
\begin{figure}
 \includegraphics[scale=0.75]{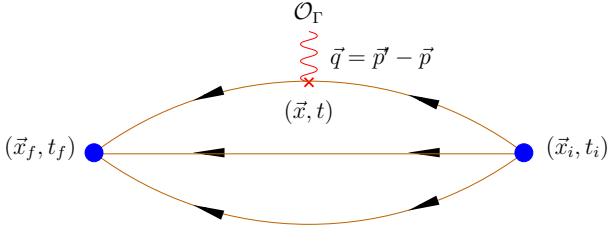}
\caption{Connected nucleon three-point function.}
\label{fig:connected_diagram}
\end{figure}

In this work we consider the nucleon matrix elements of the vector and axial-vector operators
\beq
\vspace{-1cm}
   \Op_{V^a}^{\mu_1\ldots\mu_{n}}    &=&
  \bar \psi  \gamma^{\{\mu_1}i\Dlr^{\mu_2}\ldots i\Dlr^{\mu_{n}\}}\frac{\tau^a}{2} \psi  \\
\Op_{A^a}^{\mu_1\ldots\mu_{n}}      &=&
\bar \psi  \gamma^{\{\mu_1}i\Dlr^{\mu_2}\ldots i\Dlr^{\mu_{n}\}}\gamma_5\frac{\tau^a}{2} \psi 
\label{Oper_def}\eeq
where $\tau^a$ are the Pauli matrices acting in flavor space,
$\psi$ denotes the two-component quark field (up and down). In this
work we consider the isovector combination by taking $a=3$, except
when we  discuss the spin fraction carried by each quark. Furthermore,
we limit ourselves to $n=1$ and $n=2$. The case $n=1$ reduces to the
nucleon form factors of the vector
and axial-vector currents, while $n=2$ correspond to matrix elements
of operators with a single derivative. The curly brackets represent a
symmetrization over indices and subtraction of traces, only applicable
to the operators with derivatives. There are well developed
methods to  compute the so called connected diagram, depicted in
Fig.~\ref{fig:connected_diagram}, contributing to the matrix elements
of these operators in LQCD. Each operator can be decomposed in terms
of generalized form factors (GFFs) as follows: The
matrix element of the local vector current, ${\cal O}^{\mu}_{V^3}$, is
expressed as a function of the  Dirac and Pauli form factors
\beq
&{}&\langle N(p',s')|{\cal O}_{V^3}^{\mu} |N(p,s)\rangle =\nonumber \\ 
&{}&\bar{u}_N(p',s')\left[\gamma^\mu F_1(q^2) +
  \frac{i\sigma^{\mu\nu}q_\nu}{2m_N} F_2(q^2)\right]\frac{1}{2}u_N(p,s)\nonumber\quad,
\label{Dirac ff}
\eeq
 where $u_N(p,s)$ denote the nucleon spinors of a given momentum $p$ and
spin $s$. $F_1(0)$ measures the nucleon charge while $F_2(0)$
measures the anomalous magnetic moment. They are connected to the
electric, $G_E$, and magnetic, $G_M$, Sachs form factors by the
relations \beq G_E(q^2)&=& F_1(q^2) + \frac{q^2}{(2m_N)^2}
F_2(q^2)\nonumber \\ G_M(q^2)&=& F_1(q^2) + F_2(q^2) \, .
\label{Sachs ff}
\eeq

The local axial current matrix element of the nucleon $ \langle
N(p^\prime,s^\prime) | \mathcal{O}_{A^3}^\mu | N(p,s) \rangle$ can be expressed
in terms of the form factors $ G_A$ and $G_p$ as
\beq
&&\langle N(p',s')|\mathcal{O}^\mu_{A^3}|N(p,s)\rangle= \nonumber \\
&&            \bar{u}_N(p',s') \Bigg[
            G_A(q^2)\gamma^\mu\gamma_5
            {+}\frac{q^\mu \gamma_5}{2m_N}G_p(q^2) \Bigg]\frac{1}{2}u_N(p,s)\,.\,
\label{axial ff}
\eeq

The matrix elements of the one derivative operators are
  parameterized in terms of the GFFs $A_{20}(q^2),\ B_{20}(q^2),\ C_{20}(q^2)$,
  and $\tilde A_{20}(q^2)$ and $\tilde B_{20}(q^2)$ for the vector and axial-vector operators respectively, according to
\begin{widetext}
\beq
\langle N(p^\prime,s^\prime)| \Op_{V^3}^{\mu\nu}|
N(p,s)\rangle &=&
   \bar u_N(p^\prime,s^\prime)\Bigl[ A_{20}(q^2)\, \gamma^{\{
       \mu}P^{\nu\}}
                      \hspace{-0.15cm}+B_{20}(q^2)\, \frac{i\sigma^{\{
                          \mu\alpha}q_{\alpha}P^{\nu\}}}{2m}
                      +C_{20}(q^2)\, \frac{1}{m}q^{\{\mu}q^{\nu\}} \Bigr]\frac{1}{2}
   u_N(p,s)\, ,  \\
\langle N(p^\prime,s^\prime)| \Op_{A^3}^{\mu\nu}|N(p,s)\rangle &=&
 \bar u_N(p^\prime,s^\prime)\Bigl[ \tilde A_{20}(q^2)\, \gamma^{\{
     \mu}P^{\nu\}}\gamma^5
+ \tilde B_{20}(q^2)\,
                      \frac{q^{\{\mu}P^{\nu\}}}{2m}\gamma^5\Bigr]\frac{1}{2}
 u_N(p,s)\, .\label{eq_decomposition}
\eeq
\end{widetext}
Note that the GFFs depend only on the  momentum
transfer squared, $q^2=(p^\prime-p)^2$, $p^\prime$ is the final and $p$ the initial momentum.  The isospin limit
corresponds to taking $\tau^3/2$ in Eq.~(\ref{Oper_def}) and gives the
form factor of the proton minus the form factors of the neutron. 
In the forward limit we thus have $G_E(0)=1$ and
$G_M(0)=\mu_p-\mu_n-1=4.71$~\cite{Beringer:1900zz}, which is the isovector anomalous magnetic
moment. Similarly, we obtain the nucleon axial charge, $G_A(0)\equiv
g_A$, the isovector momentum fraction, $A_{20}(0) \equiv \langle x
\rangle_{u-d} $ and the moment of the polarized quark distribution,
$\tilde A_{20}(0) \equiv \langle x \rangle_{\Delta u-\Delta d} $.  In
order, to find the spin and angular momentum 
carried by each quark individually in the nucleon we need  the
isoscalar  axial charge and the isoscalar one-derivative matrix elements
of the vector operator. Unlike the isovector combinations,
where disconnected fermion loops vanish in the continuum limit, the
isoscalar cases receive contributions from disconnected fermion
loops. The evaluation of the disconnected 
contributions is difficult due to the computational cost but
techniques are being developed to compute them. Recent 
results on nucleon form factors show that they small or consistent
with zero~\cite{Alexandrou:2012zz, Babich:2010at,
  Alexandrou:2012py}. The disconnected contribution to the isoscalar
axial charge has been contributed and was found to be nonzero, but it is an
order of magnitude smaller than the connected one~\cite{QCDSF:2011aa}. 
Therefore in most nucleon structure calculations they are neglected.
In this work we 
will  assume that the disconnected contributions are small, in which case, it is straightforward to evaluate the 
isoscalar matrix elements taking into account only the connected part depicted in Fig.~\ref{fig:connected_diagram}. The quark contribution to the nucleon spin
is obtained using Ji's sum rule: $J^q = \frac{1}{2}[ A_{20}^q(0) +
  B_{20}^q(0)]$. Moreover, using the axial charge for each quark, $g_A^q$, 
we obtain the intrinsic spin of each quark, $\Delta \Sigma^q=g_A^q$,  and via 
the decomposition $J^q=\frac{1}{2}\Delta \Sigma^q+ L^q$ we can extract
the quark orbital angular momentum $L^q$.

\medskip
In the present work we employ the twisted mass
fermion (TMF) action~\cite{Frezzotti:2000nk} and the Iwasaki improved
gauge action~\cite{Weisz:1982zw}. Twisted mass fermions provide an
attractive formulation of lattice QCD that allows for automatic ${\cal
O}(a)$ improvement, infrared regularization of small eigenvalues and
fast dynamical simulations~\cite{Frezzotti:2003ni}. In the computation of 
GFFs the automatic ${\cal O}(a)$ improvement is particularly relevant
since it is achieved by tuning only one parameter in the action,
requiring no further improvements on the operator level.

We use the twisted mass Wilson action for the light doublet of quarks
\begin{equation}
S_l = \sum_x \bar\chi_l(x) \left[D_W {+} m_{(0,l)} {+} i\gamma_5\tau^3\mu_l \right]\chi_l(x)\,,
\end{equation}
where $D_W$ is the Wilson Dirac operator, $m_{(0,l)}$ is the
untwisted bare quark mass, $\mu_l$ is the bare light twisted mass. The quark
fields $\chi_l$ are in the so-called ``twisted basis'' obtained from the
``physical basis'' at maximal twist by the transformation
\be
\psi {=} \frac{1}{\sqrt{2}}[{\bf 1} + i\tau^3\gamma_5]\chi_l \quad {\rm and}
\quad \bar\psi {=} \bar\chi_l \frac{1}{\sqrt{2}}[{\bf 1} + i\tau^3\gamma_5]\,.
\ee
In addition to the light sector, we introduce a twisted heavy
mass-split doublet $\chi_h = (\chi_c,\chi_s)$ for the strange and
charm quarks, described by the action
\begin{equation}
S_h = \sum_x \bar\chi_h(x) \left[D_W {+} m_{(0,h)} {+}
  i\gamma_5\tau^1\mu_\sigma + \tau^3\mu_\delta\right]\chi_h(x) \>,
\end{equation}
where $m_{(0,h)}$ is the untwisted bare quark mass for the heavy
doublet, $\mu_\sigma$ is the bare twisted mass along the $\tau^1$
direction and $\mu_\delta$ is the mass splitting in the $\tau^3$ direction. 
The quark mass $m_{(0,h)}$ is set equal to $m_{(0,l)}$ in the simulations
thus ensuring ${\cal O}(a)$-improvement also in the heavy quark sector.
The chiral rotation for the heavy quarks from the twisted to the
physical basis is
\be
\psi {=} \frac{1}{\sqrt{2}}[{\bf 1} + i\tau^1\gamma_5]\chi_h \quad {\rm and}
\quad \bar\psi {=} \bar\chi_h \frac{1}{\sqrt{2}}[{\bf 1} + i\tau^1\gamma_5]\,.
\ee
The reader can find more details on the twisted mass fermion action in Ref.~\cite{Baron:2010bv}.
Simulating a charm quark may give rise to concerns regarding cut-off effects.
The  observables of this work cannot be used to check for such effect. However,
an analysis in Ref~\cite{Athenodorou:2011zp}
shows that they are surprising small.

\subsection{Correlation functions}

The  GFFs are extracted from dimensionless ratios of
correlation functions, involving two-point and three-point functions
that are defined by
\begin{widetext}
\be
G(\vec q, t_f-t_i)=
\sum_{\vec x_f} \, e^{-i(\vec x_f-\vec x_i) \cdot \vec q}\,
     {\Gamma^0_{\beta\alpha}}\, \langle {J_{\alpha}(t_f,\vec x_f)}{\overline{J}_{\beta}(t_i,\vec{x}_i)} \rangle \label{twop}
\ee
\beq
G^{\mu_1 \cdots\mu_n}(\Gamma^\nu,\vec q,t) =
\sum_{\vec x, \vec x_f}e^{i(\vec x -\vec x_i)\cdot \vec q}\,  \Gamma^\nu_{\beta\alpha}\, \langle
{J_{\alpha}(t_f,\vec x_f)} \Op^{\mu_1 \cdots \mu_n}(t,\vec x) {\overline{J}_{\beta}(t_i,\vec{x}_i)}\rangle \>.\label{threep}
\eeq
\end{widetext}
For the insertion, $\Op^{\mu_1 \cdots \mu_n}$, we employ the
vector ($\bar \psi\,\gamma^{\mu}\psi$), the axial-vector ($\bar
\psi\,\gamma^5\,\gamma^{\mu}\psi$), the one-derivative vector ($\bar
\psi\,\gamma^{\{\mu_1} D^{\mu_2\}}\psi$) and the one-derivative axial-vector ($\bar
\psi\,\gamma5\,\gamma^{\{\mu_1} D^{\mu_2\}}\psi$) operators.
We consider kinematics for which the final momentum $\vec{p}^\prime=0$ and in our
approach we fix the time separation between sink and source $t_f-t_i$. The projection matrices
${\Gamma^0}$ and ${\Gamma^k}$ are given by
\be
{\Gamma^0} = \frac{1}{4}(\eins + \gamma_0)\,,\quad {\Gamma^k} =
{\Gamma^0} i \gamma_5  \gamma_k \, .\label{proj}
\ee
 The proton interpolating field written in terms of the quark fields
 in the twisted basis ($\tilde{u}$ and $\tilde{d}$) at maximal twist
 is given by
\be
{J}(x) {=} {\frac{1}{\sqrt{2}}[\eins + i\gamma_5]}\epsilon^{abc} \left[ {\tilde{u}}^{a \top}(x) \C\gamma_5 \tilde{d}^b(x)\right] {\tilde{u}}^c(x)\,,
\ee
where $\C$ is the charge conjugation matrix. We use Gaussian smeared
quark fields~\cite{Alexandrou:1992ti,Gusken:1989} to increase
the overlap with the proton state and decrease overlap with excited
states. The smeared interpolating fields are given by
\beq
q_{\rm smear}^a(t,\vec x) &=& \sum_{\vec y} F^{ab}(\vec x,\vec y;U(t))\ q^b(t,\vec y)\,,\\
F &=& (\eins + {a_G} H)^{N_G} \,, \nonumber\\
H(\vec x,\vec y; U(t)) &=& \sum_{i=1}^3[U_i(x) \delta_{x,y-\hat\imath} + U_i^\dagger(x-\hat\imath) \delta_{x,y+\hat\imath}]\,. \nonumber
\eeq
We also apply APE-smearing to the gauge fields $U_\mu$ entering
the hopping matrix $H$. The parameters for the Gaussian smearing
$a_G$ and $N_G$ are optimized using the nucleon ground state~\cite{Alexandrou:2008tn}.
Different combination of Gaussian parameters, $N_{\rm G}$ and $a_{\rm G}$, have
been tested and it was found that combinations of $N_G$ and $a_G$ that
give a root mean square radius of about 0.5~fm  are optimal for suppressing
excited states. The
results of this work have been produced with 
\beq
\beta=1.95: & N_G=50\>,\, a_G=4,\,N_{\rm APE}=20,\,a_{\rm APE}=0.5,&\nonumber \\
\beta=2.10: & N_G=110,\, a_G=4,\,N_{\rm APE}=50,\,a_{\rm APE}=0.5\,. &\nonumber
\eeq
As already point out, in correlators of isovector operators the disconnected
diagrams are zero up to lattice artifacts, and can be safely neglected
as we approach the continuum limit. Thus, these correlators can be calculated
by evaluating the connected diagram of Fig.~\ref{fig:connected_diagram}
for which we employ sequential inversions through the sink~\cite{Dolgov:2002zm}.
The creation operator is taken at a fixed position $\vec{x}_i {=}
{\vec 0}$ (source). The annihilation operator at a later time $t_f$
(sink) carries momentum $\vec{p}^\prime {=} 0$. The current couples to
a quark at an intermediate time $t$ and carries momentum $\vec{q}$. Translation invariance enforces $\vec{q}=-\vec{p}$ for our
kinematics. At a fixed sink-source time separation we obtain results
for all possible momentum transfers and insertion times as well as for
any operator $\Op_\Gamma^{\{\mu_1\cdots\mu_n\}}$, with one set of sequential
inversions per choice of the sink. We perform separate inversions for
the two projection matrices $\Gamma^0$ and $\sum_k \Gamma^k$   given in Eq.~(\ref{proj}).
An alternative approach that computes the spatial all-to-all propagator 
stochastically  has shown ot be suitable for the evaluation of nucleon 
three-point functions~\cite{Alexandrou:2013xon}. Within this approach
 one can include any projection without needing additional inversions.

\begin{figure}[h]
   \includegraphics[scale=0.3,angle=-90]{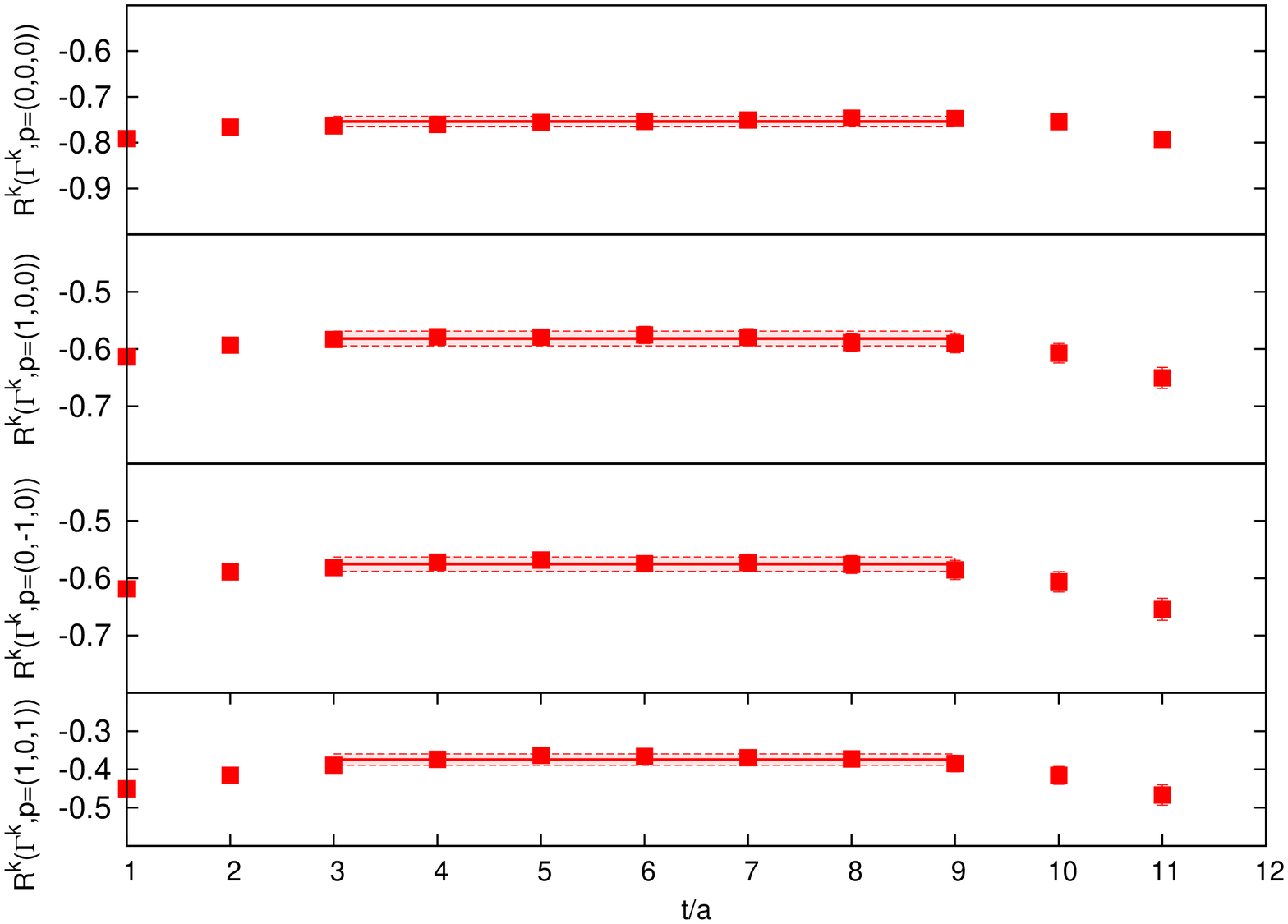}
\includegraphics[scale=0.3,angle=-90]{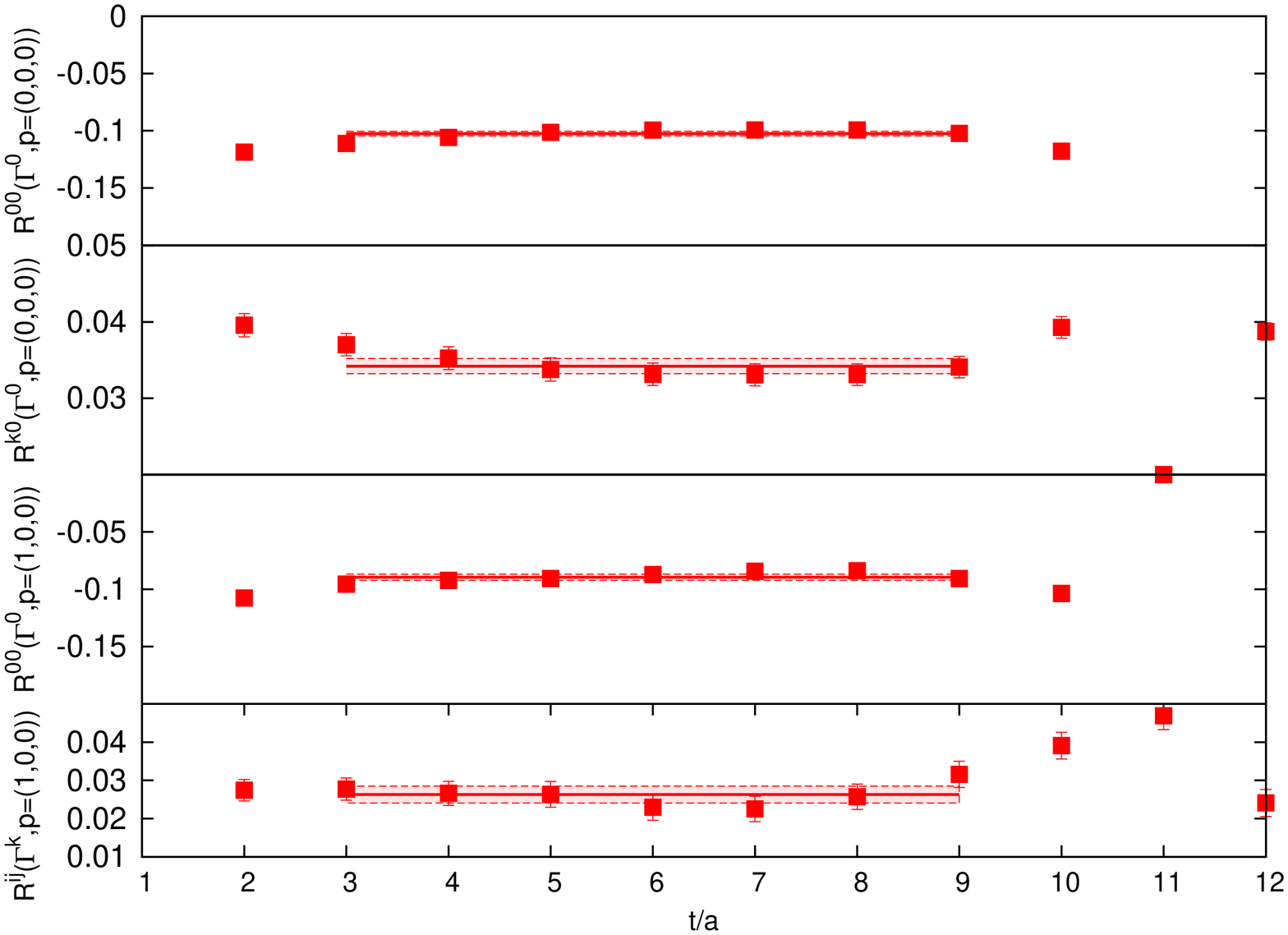}
   \caption{\label{fig:plateaus} Ratios for the matrix elements of the local axial-vector operator (upper) and
     one derivative vector operator (lower) for a few exemplary
     choices of the momentum. The solid lines with the bands indicate
     the fitted plateau values with their jackknife errors. From top
     to bottom the momentum takes values $\vec{p}=(0,0,0),\,(1,0,0),\,
     (0,-1,0)$ and $(1,0,1)$.}
\end{figure}

Using the two- and three-point functions of 
Eqs.~(\ref{twop})-(\ref{threep}) and considering operators with up to
one derivative we form the ratio 
\beq
R^{\mu\nu}(\Gamma^\lambda,\vec q,t)&=& \frac{G^{\mu\nu}(\Gamma^\lambda,\vec q,t) }{G(\vec 0,
  t_f-t_i)} \nonumber\\
&{}& \hspace*{-1.2cm}\times\sqrt{\frac{G(\vec p, t_f{-}t)G(\vec 0,  t-t_i)G(\vec0,
    t_f-t_i)}{G(\vec 0  , t_f{-}t)G(\vec p,t-t_i)G(\vec p,t_f-t_i)}}\>,
\label{ratio}
\eeq
which is optimized because it does not contain potentially noisy
two-point functions at large separations and because correlations
between its different factors reduce the statistical noise.
For sufficiently large separations $t_f-t$ and $t-t_i$ this ratio
becomes time-independent (plateau region):
\be
\lim_{t_f-t\rightarrow \infty}\lim_{t-t_i\rightarrow
  \infty}R^{\mu\nu}(\Gamma^\lambda,\vec
q,t)=\Pi^{\mu\nu}(\Gamma^\lambda,\vec q) \,.
\label{plateau}
\ee
From the plateau values of the renormalized asymptotic ratio
$\Pi(\Gamma^\lambda, \vec q)_R=Z\,\Pi(\Gamma^\lambda, \vec{q})$ 
the nucleon matrix elements of all 
our operators can be extracted. The equations
relating $\Pi(\Gamma^\lambda,\vec q)$ to the GFFs can be found in 
Refs.~\cite{Alexandrou:2011nr,Alexandrou:2011db,Alexandrou:2010hf}.
All values of $\vec q$ resulting in the same $q^2$, the two choices
of projector matrices $\Gamma^0$ and $\sum_k \Gamma^k$ given Eq.~(\ref{proj})
 and the relevant orientations $\mu,\nu$ of the operators
lead to an over-constrained system of equations, which is solved in the
least-squares sense via a singular value decomposition of the
coefficient matrix. All quantities will be given in Euclidean space
with $Q^2\equiv-q^2$ the Euclidean momentum transfer squared.
Both projectors $\Gamma^0$ and $\sum_k\Gamma^k$ are required to obtain all
 GFFs, except for the case of the local axial-vector
operator, for which the projection with $\Gamma^0$ leads to zero.
For the one derivative vector operator, both cases $\mu=\nu$ and
$\mu\neq \nu$ are necessary to extract all three GFFs, which on a
lattice renormalize differently from each
other~\cite{Gockeler:1996mu}. On the other hand, the one-derivative
axial-vector form factors can be extracted using only correlators with
$\mu\neq \nu$, but we use all combinations of $\mu,\nu$ in order to
increase statistics. In Fig.~\ref{fig:plateaus} we show
representative plateaus for the ratios of the local axial-vector and the one
derivative vector operators at $\beta=1.95$, using different momenta, projectors,
and indices $\mu,\,\nu$.

Since we use sequential inversions through the sink we need to fix the
sink-source separation. Optimally, one wants to keep 
 the statistical errors on the ratio of Eq.~(\ref{ratio})  as small as possible by using the smallest value
for 
the sink-source time separation that still ensures that the
excited state contributions are sufficiently suppressed. 
Recent studies have shown that the optimal sink-source separation
is operator dependent~\cite{Alexandrou:2011aa,Dinter:2011sg}. For $g_A$ excited
state contamination was found to be small.  
We have also tested different values of the sink-source time
separation~\cite{Alexandrou:2010hf} for the magnetic form factor
and found consistent results when the sink-source separation was about 1~fm
within our statistical accuracy. For the momentum fraction
one would need to re-examine the optimal sink-source separation, which would require a dedicate high accuracy study. 
 Since in this work we are computing several observables, we will  
   use $t_f-t_i\sim 1$ fm that correspond to  the following values
\be
\beta=1.95:\,(t_f-t_i)/a{=}12\,,\quad\beta=2.10:\,(t_f-t_i)/a{=}18. \nonumber
\ee
This choice allows to compare with other lattice QCD results where similar values were used.

\subsection{Simulation details}

In Table~\ref{Table:params} we tabulate the input parameters of the
calculation, namely $\beta$, $L/a$ and the light quark mass $a\mu$, as well as, the value of the pion
mass in lattice units~\cite{Baron:2010bv,Baron:2011sf}. The strange and charm quark masses were fixed to
 approximately reproduce the physical kaon and D-meson masses, respectively~\cite{Baron:2010th}. 
 The lattice spacing $a$  given in this Table is determined from the nucleon
mass as explained in the following subsection and it will be used for  the baryon observables discussed in this paper. We note that
the study of the systematic error in the  scale setting using the pion decay constant as compared to the value extracted using the nucleon mass is currently being pursued.
Since the GFFs are dimensionless they are  not affected by the 
scale setting. However, $a$ is needed to convert $Q^2$ to physical units,
and therefore it does affect quantities like the anomalous magnetic moment
and Dirac and Pauli radii since these are dimensionful parameters that
depend on fitting the $Q^2$-dependence of the form factors.

\begin{center}
\begin{table}[h]
\begin{tabular}{c|llll}
\hline\hline
\multicolumn{3}{c}{$\beta=1.95$, $a=0.0820(10)$~fm,   ${r_0/a}=5.66(3)$}\\\hline
$32^3\times 64$, $L=2.6$~fm  &$a\mu$ & 0.0055  \\
                               & No. of confs & 950  \\
                               & $a\,m_\pi$& 0.15518(21)(33)\\
                               & $Lm_\pi $    & 4.97   \\\hline \hline
%
\multicolumn{3}{c}{ $\beta=2.10$, $a=0.0644(7)$~fm, ${r_0/a}=7.61(6)$ }\\
\hline
$48^3\times 96$, $L=3.1$~fm &$a\mu$         & 0.0015   \\
                               & No. of confs   &900 \\
                               &$a\,m_\pi$ & 0.06975(20)  \\
                               &$Lm_\pi$     & 3.35         \\ \hline
\end{tabular}
\caption{Input parameters ($\beta,L,a\mu$) of our lattice calculation
  with the corresponding lattice spacing $a$, determined from the
  nucleon mass, and pion mass $a m_{\pi}$ in lattice units.}
\label{Table:params}
\end{table}
\end{center}

\subsection{Determination of lattice spacing}

For the observables discussed in this work the
nucleon mass at the physical point is the most appropriate quantity to
set the scale. The values for the nucleon mass were
computed using $N_f{=}2{+}1{+}1$ ensembles for  $\beta{=}1.90$,
$\beta{=}1.95$ and $\beta{=}2.10$, a range of pion masses and volumes. 
To extract the mass we consider the two-point correlators defined in
Eq.~(\ref{twop}) and construct the effective mass
\beq
a\,m^{\rm eff}_N(t) \hspace{-0.2cm}&=&\hspace{-0.2cm} -\log(C(t)/C(t-1)) \nonumber \\
\hspace{-0.2cm}&=&\hspace{-0.2cm}a\, m_N + \log(\frac{1+\sum_{j=1}^\infty c_j e^{-\Delta_j t}}
{1+\sum_{j=1}^\infty c_j e^{-\Delta_j (t-1)}})\nonumber \\
\hspace{-0.2cm}&\xrightarrow{t\rightarrow \infty}\, a\,m_N  
\label{mN}
\eeq
where $\Delta_j= E_j-m_N$ is the energy difference of the excited state
$j$ with respect to the ground state mass, $m_N$.
Our fitting procedure to extract $m_N$ is as follows: The mass is
obtained from  a constant
fit to $m^{\rm eff}_N(t)$ for  $t\ge t_1$ for which 
the contamination of excited states is believed to be small. We denote
the value extracted as $m_N^{(A)}(t_1)$. A second fit  to $m^{\rm
  eff}_N(t)$ is performed including the first excited state for $t\ge
t_1^\prime$, where $t_1^\prime$ is taken to be $2a$ or $3a$.  We
denote the  value for the ground state mass extracted from the fit to
two exponentials by $m_N^{(B)}$. We vary $t_1$  such that the ratio
\beq
\frac{|a\,m_N^{(A)}(t_1) -a\,m_N^{(B)}|}{a\,m_N^{mean}},\quad
     {\rm where} \nonumber \\
a\,m_N^{mean} = \frac{a\,m_N^{(A)}(t_1) + a\,m_N^{(B)}}{2}
\eeq
drops below 50\% of the statistical error on $m_N^A(t_1)$.
The resulting values for the nucleon mass are collected in
Table~\ref{tab:mN}. 

\begin{table}[h]
\begin{center}
\begin{tabular}{||c||c||c||c||c||c||}
\hline
\hline
$\,\,\,$$\beta$$\,\,\,$  & $\,\,\,$ $a\,\mu$$\,\,\,$   &  $\,\,\,$Volume$\,\,\,$ &  $\,\,\,am_\pi\,\,\,$ & statistics & $\,\,\,$$am_N$$\,\,\,$   \\
\hline
1.90  &    0.003   &  32$^3\times$64   &0.124  &740        & 0.524(9)  \\
1.90  &    0.004   &  20$^3\times$48   &0.149  &617        & 0.550(19)  \\
1.90  &    0.004   &  24$^3\times$48   &0.145  &2092       & 0.541(8)  \\
1.90  &    0.004   &  32$^3\times$64   &0.141  &1556       & 0.519(11)  \\
1.90  &    0.005   &  32$^3\times$64   &0.158  &387        & 0.542(6)  \\
1.90  &    0.006   &  24$^3\times$48   &0.173  &1916       & 0.572(5)  \\
1.90  &    0.008   &  24$^3\times$48   &0.199  &1796       & 0.590(5)  \\
1.90  &    0.010   &  24$^3\times$48   &0.223  &2004       & 0.621(4)  \\
\hline
1.95  &    0.0025   &  32$^3\times$64   &0.107  & 2892     & 0.447(6)  \\
1.95  &    0.0035   &  32$^3\times$64   &0.126  &4204      & 0.478(5)  \\
1.95  &    0.0055   &  32$^3\times$64   &0.155  &18576     & 0.503(2)  \\
1.95  &    0.0075   &  32$^3\times$64   &0.180  &2084      & 0.533(4)  \\
1.95  &    0.0085   &  24$^3\times$48   &0.194  & 937      & 0.542(5)  \\
\hline
2.10  &    0.0015   &  48$^3\times$96   &0.070  &2424     &  0.338(4)  \\
2.10  &    0.0020   &  48$^3\times$96   &0.080  & 744     &  0.351(7)  \\
2.10  &    0.0030   &  48$^3\times$96   &0.098  & 226     &  0.362(7)  \\
2.10  &    0.0045   &  32$^3\times$64   &0.121  &1905     &  0.394(3)  \\
\hline\hline
\end{tabular}
\caption{Values of the nucleon mass and the associated  statistical error.}
\label{tab:mN}
\end{center}
\end{table}

\begin{figure}
\includegraphics[scale=0.28,angle=-90]{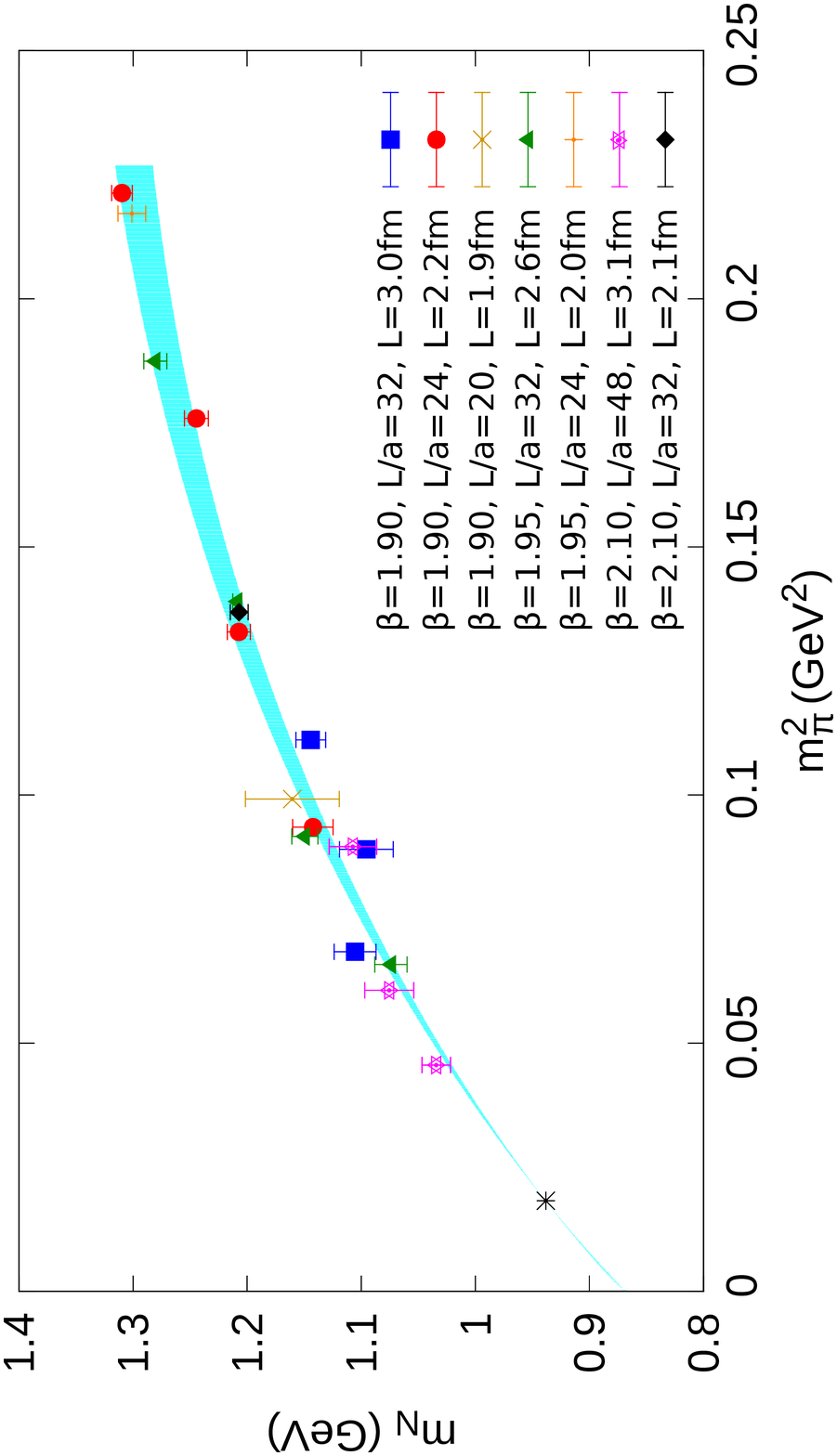}\\\vspace*{0.3cm}
\includegraphics[scale=0.28,angle=-90]{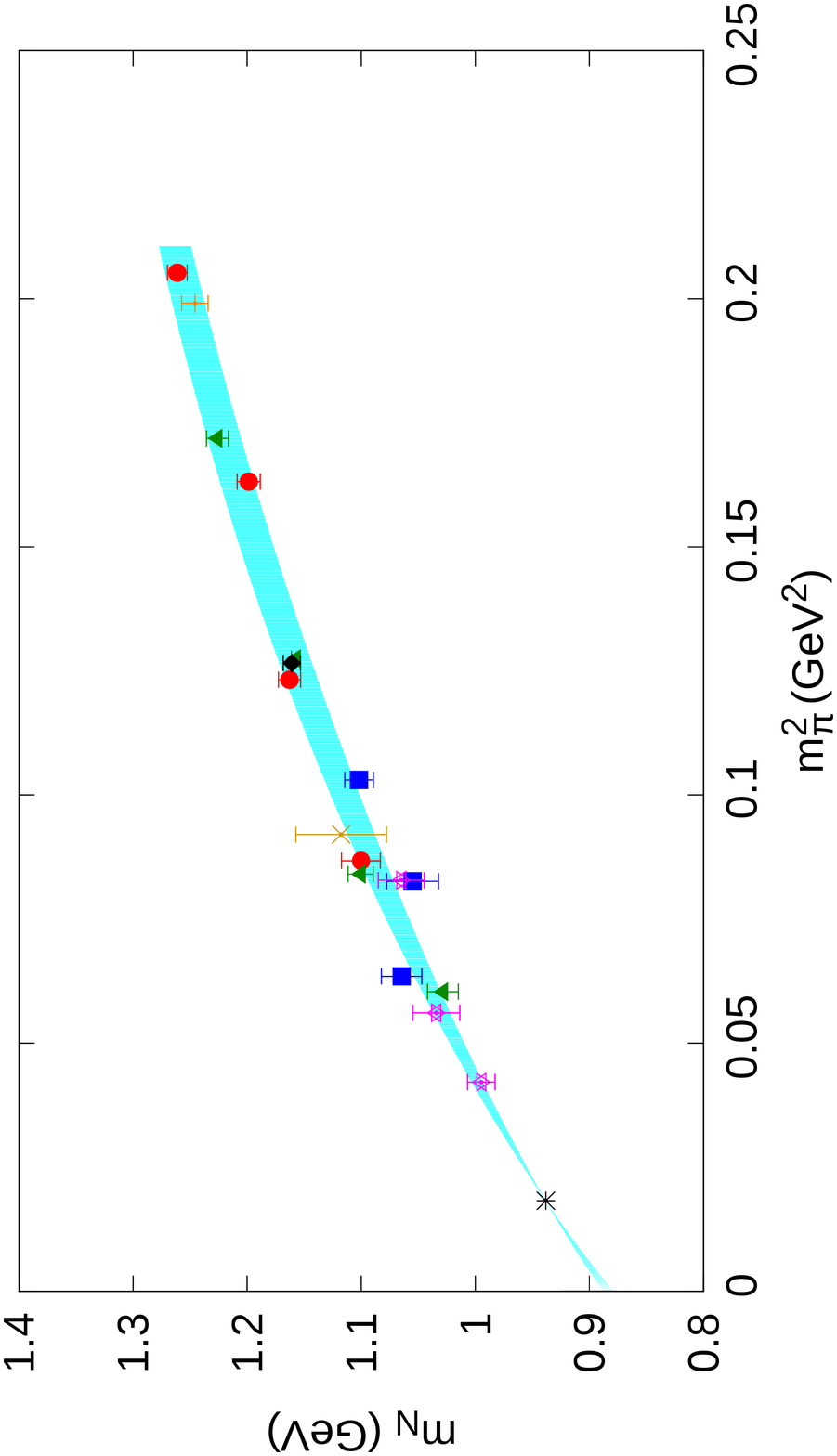}\\
\caption{Nucleon mass at three lattice spacings. The solid lines
are fits to ${\cal O}(p^3)$ (upper panel) and ${\cal O}(p^4)$ (lower panel) HB$\chi$PT with explicit $\Delta$ degrees of freedom in the so called small scale expansion(SSE).  The dotted lines
denote the error band. The
physical point is shown with the asterisk.}
\label{fig:nucleon scale}
\end{figure}

In Fig.~\ref{fig:nucleon scale} we show results at three values of the lattice
spacing corresponding to $\beta{=}1.90$, $\beta{=}1.95$ and
$\beta{=}2.10$. As can be seen, cut-off effects are negligible and we
can therefore use continuum chiral perturbation theory to extrapolate to
the physical point using all the lattice results. 

To chirally extrapolate we use the well-established ${\cal O}(p^3)$ result of
 chiral perturbation theory ($\chi$PT) given by
\be  
m_N = {m_N^0}-{4c_1}m_\pi^2 -\frac{3 g_A^2 }{16\pi f_\pi^2} m_\pi^3\,.
\label{p3}
\ee
We perform a fit to the results at the three $\beta$
values given in Table~\ref{tab:mN} using the  ${\cal O}(p^3)$ 
expansion of Eq.~(\ref{p3}) with fit parameters $m_N^0$, $c_1$ and 
the three lattice spacings. The resulting fit is shown in Fig.~\ref{fig:nucleon scale} and describes well our lattice data ($\chi^2/$d.0.f) yielding 
for the lattice spacings the values 
\beq
a_{\beta=1.90}&=&0.0934(13)(35)~{\rm fm}\,, \nonumber\\
a_{\beta=1.95}&=&0.0820(10)(36)~{\rm fm}\,,\nonumber\\
a_{\beta=2.10}&=&0.0644(7)(25)~{\rm fm}\,.
\label{lattice spacings}
\eeq
We would like to point out that our lattice results show a curvature supporting
the $m_\pi^3$-term. 
In order to estimate the systematic error due to the
chiral extrapolation we also perform a fit using heavy baryon (HB) $\chi$PT to  ${\cal O}(p^4)$ with explicit $\Delta$ degrees of freedom in the so called small scale expansion(SSE)~\cite{Alexandrou:2008tn}.
 We take the difference between the
${\cal O}(p^3)$ and ${\cal O}(p^4)$ mean values as an estimate of the
uncertainty due to the chiral extrapolation. This error is given in the
second parenthesis in Eqs.~(\ref{lattice spacings}) and it is about twice the
statistical error. 
In order to assess discretization errors we perform a fit to ${\cal
  O}(p^3)$ at each value of $\beta$ separately. We find
$a=0.0920(21),\> 0.0818(16),\> 0.0655(12)$~fm at $\beta=1.90,\>
1.95,\>2.10$ respectively. These values are fully consistent with
those obtained in Eq.~(\ref{lattice spacings}) indicating that
discretization effects are small confirming a posteriori the validity
of assuming that cut-off effects are small. 
The values of the lattice spacing given in Eqs.~(\ref{lattice
  spacings}) will be used for converting to physical units the
quantities we study here. 
We would like to point out that redoing the ${\cal O}(p^3)$ fit
eliminating data for which $L m_\pi< 3.5$ yields $a_{\beta=1.90}=0.0942(14)$~fm, $a_{\beta=1.95}= 0.0858(11)$~fm and $a_{\beta=2.1}=0.0653(8)$, which are
consistent with the values given in Eq.~(\ref{lattice spacings}).
In performing these fits we only take
into account statistical errors. Systematic errors due to the choice of the plateau are not included. 
We also note that the lattice spacings were also
determined from the pion decay constant using NLO SU(2) chiral perturbation theory to extrapolate the lattice data. The values  obtained at $\beta=1.90$, 1.95 and 2.10 in this preliminary analysis that included only a subset of 
the ensembles used here
 are smaller~\cite{Baron:2011sf},  as compared 
to the values  extracted using the nucleon mass. For the two $\beta$-values
studied in this work they were found to be 
$a_{f_\pi}=0.0779(4)$~fm at $\beta=1.95$ and $a_{f_\pi}=0.0607(3)$~fm at $\beta=2.10$, where  with $a_{f_\pi}$ we denote the lattice spacing determined using the pion decay constant. This means
that the values of the pion mass in physical units quoted in this paper are equivalently 
smaller than those obtained using $a_{f_\pi}$ to convert to physical units. 
A comprehensive analysis  of the scale setting and the associated
systematic uncertainties is currently being carried out  by European Twisted Mass Collaboration~(ETMC) and will appear elsewhere.

\subsection{Renormalization}
\begin{figure}
\includegraphics[scale=0.35]{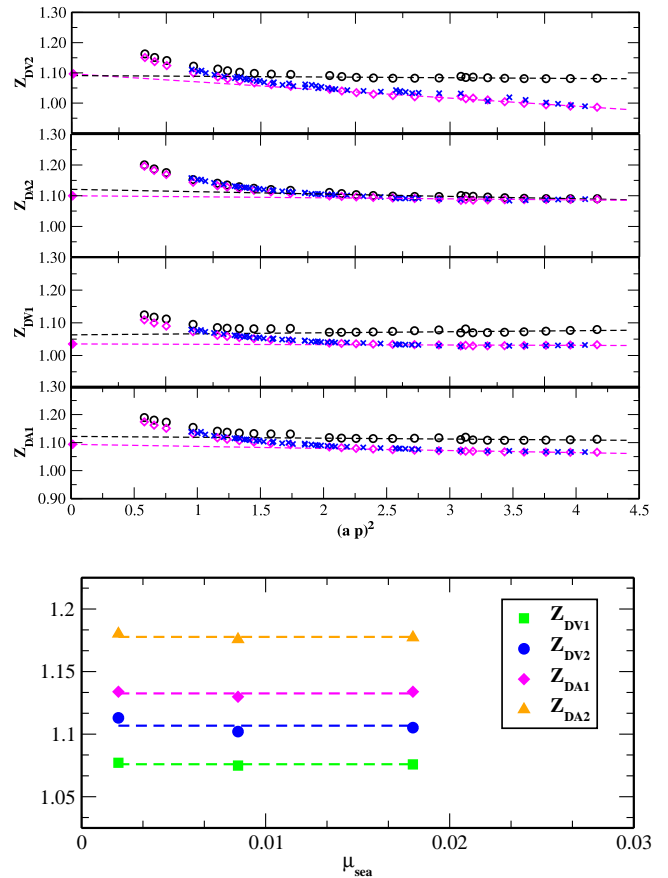}\\[3ex]
$\qquad$\includegraphics[scale=0.45]{plots/Z_oneD_b1.95_mpi_dependence_theta_aver.eps}
\caption{Upper panel: One derivative renormalization functions for
  $\beta=2.10,\,a\,\mu=0.0015$ using $N_f{=}4$ gauge
  configurations, where $Z_{DV1}\, (Z_{DA1}) \equiv Z_{DV}^{\mu\mu}\, (Z_{DA}^{\mu\mu})$ 
and $Z_{DV2}\, (Z_{DA2}) \equiv Z_{DV}^{\mu\neq\nu}\, (Z_{DA}^{\mu\neq\nu})$.
Black circles are the unsubtracted data and the
  magenta diamonds the data after subtracting the perturbative ${\cal
    O}(a^2)$-terms. For comparison, we show the subtracted data using
  $N_f{=}2{+}1{+}1$ gauge configurations at the same value of the
  quark mass and $\beta$ (blue crosses). Lower panel: One derivative
  renormalization functions for $\beta=1.95$ using $N_f{=}4$ gauge
  configurations as a function of the twisted quark mass.}
\label{Z_Nf4_Nf211}
\end{figure}

We determine the renormalization constants needed for
the operators discussed in this work in the RI$'$-MOM
scheme~\cite{Martinelli:1994ty} 
by employing a momentum source at the vertex~\cite{Gockeler:1998ye}.
The advantage of this method is the high statistical accuracy and the
evaluation of the vertex for any operator including extended operators
at no significant additional computational cost. For the details of
the non-perturbative renormalization see Ref.~\cite{Alexandrou:2010me}.
In the RI scheme the renormalization constants are defined in the
chiral limit. Since the mass of the strange and charm quarks are fixed
to their physical values in these simulations, extrapolation to the
chiral limit is not possible. Therefore, in order to compute the
renormalization constants needed to obtain physical observables, ETMC
has generated $N_f{=}4$ ensembles for the same $\beta$ values so that the
chiral limit can be taken~\cite{Dimopoulos:2011wz}. 
Although we will use the $N_f{=}4$ ensembles for the final determination
of the renormalization constants, it is also interesting to compute the
renormalization constants using the $N_f{=}2{+}1{+}1$ ensembles and study
their quark mass dependence.
This test was performed on both the $\beta=1.95$ and the $\beta=2.10$ ensembles. In
the upper panel of Fig.~\ref{Z_Nf4_Nf211} we show results at
$\beta=2.10$ for both $N_f{=}4$ and $N_f{=}2{+}1{+}1$ ensembles for the one
derivative Z-factors in the RI$'$-MOM scheme. As can be seen, we
obtain compatible values for all four cases. We also observe the same
agreement for $Z_V$ and $Z_A$ also  at $\beta=1.95$. This can be
understood by examining the quark mass dependence of these renormalization constants. In
the lower panel of Fig.~\ref{Z_Nf4_Nf211} we show, for the $N_f{=}4$
case, the dependence of $Z_{DV}$, $Z_{DA}$ on four light quark
masses. The values we find are consistent with each other. This
explains the fact that the results in the $N_f{=}4$ and $N_f{=}2{+}1{+}1$
cases are compatible. Furthermore, it makes any extrapolation of
$N_f{=}4$ results to the chiral limit straight forward.
\begin{figure}[h]
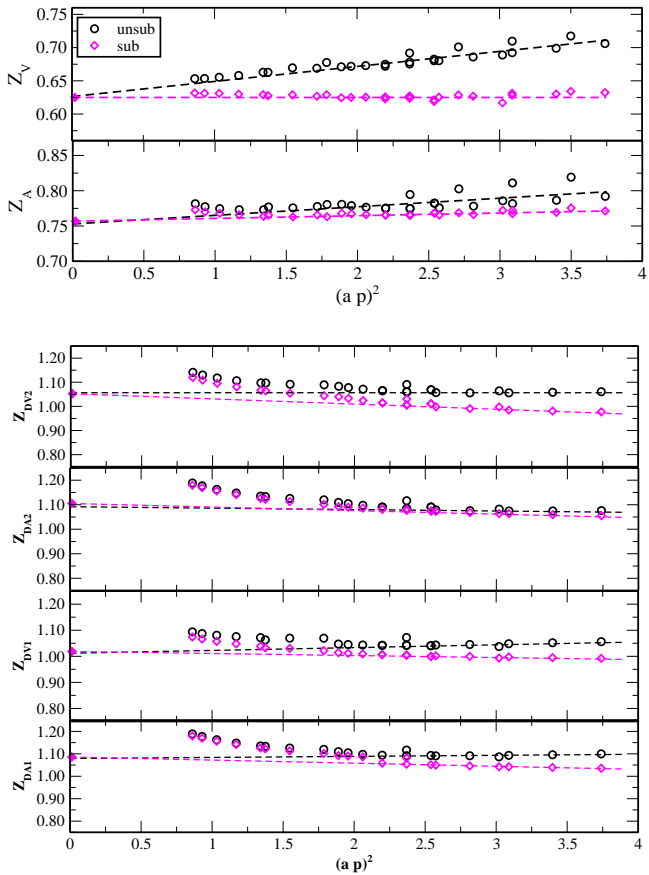

\includegraphics[scale=0.35]{plots/Z_v_a_b1.95.eps}\\vspace*{0.5cm}
\includegraphics[scale=0.35]{plots/Z_oneD_MSbar_2GeV_b1.95.eps}
\caption{Upper panel: $Z_A,\,Z_V$ for $\beta=1.95,$ and $a\mu=0.0055$;
  Lower panel:  Renormalization constants for one derivative operators
  for $\beta=1.95,$ and $a\mu=0.0055$, where 
$Z_{DV1}\, (Z_{DA1}) \equiv Z_{DV}^{\mu\mu}\, (Z_{DA}^{\mu\mu})$
and $Z_{DV2}\, (Z_{DA2}) \equiv Z_{DV}^{\mu\neq\nu}\,(Z_{DA}^{\mu\neq\nu})$.
The lattice data are shown in black circles and the data after the
${\cal O}(a^2)$-terms have been subtracted are shown in magenta
diamonds. The solid diamond at $(a\,p)^2=0$ is the value obtained
after performing a linear extrapolation of the subtracted data.}
\label{Z_oneD}
\end{figure}
We perform a perturbative subtraction of ${\cal O}(a^2)$-terms~\cite{Constantinou:2009tr,Alexandrou:2010me,Alexandrou:2012mt}.
This subtracts the leading cut-off effects yielding, in general, a  weak
dependence of the renormalization factors on $(ap)^2$ for which the
$(ap)^2\rightarrow 0$ limit can be reliably taken, as can be seen in Figs.~\ref{Z_Nf4_Nf211} and \ref{Z_oneD} for the two $N_f=2+1+1$ ensembles. We also take the
chiral limit, although the quark mass dependence is negligible for
the aforementioned operators.

The renormalization factors for the one-derivative vector and
axial-vector operators, $Z_{DV}^{\mu\nu}$ and $Z_{DA}^{\mu\nu}$, fall
into different irreducible representations of the hypercubic group,
depending on the choice of the external indices, $\mu$, $\nu$. Hence,
we distinguish between $Z_{DV}^{\mu\mu}\, (Z_{DA}^{\mu\mu})$ and 
$Z_{DV}^{\mu\neq\nu}\, (Z_{DA}^{\mu\neq\nu})$. For the conversion factors from RI to 
$\overline{\rm MS}$ we used the results of Ref.~\cite{Gracey:2003yr} for the local 
vector and axial-vector operators while for the one-derivative operators
we used the expressions of Ref.~\cite{Alexandrou:2010me}. 
Another characteristic of these renormalization constants is that they
depend on the renormalization scale. Thus, they need to be converted
to the continuum $\overline{\rm MS}$-scheme, and for this we use a
conversion factor computed in perturbation theory to $O(g^4)$. They 
are also evolved perturbatively to a reference scale, which is chosen
to be (2~GeV)$^2$. The results are shown in Fig.~\ref{Z_oneD} both before
subtracting the perturbative ${\cal O}(a^2)$-terms and after. Using
the subtracted data we find the values given in
Table~III.

\begin{table}[h]
\begin{center}
\begin{tabular}{|c|cc|}
\hline
 & $\beta{=}$1.95 & $\beta{=}$2.10\\
\hline
$Z_V$ & 0.625(2) &  0.664(1)  \\
$Z_A$ & 0.757(3) &  0.771(2)  \\
$Z_{DV}^{\mu\mu}$ & 1.019(4) & 1.048(5) \\
$Z_{DV}^{\mu\ne\nu}$&1.053(11) & 1.105(4) \\
$Z_{DA}^{\mu\mu}$ & 1.086(3) & 1.112(5)  \\ 
$Z_{DA}^{\mu\ne\nu}$& 1.105(2) & 1.119(6)\\
\hline
\end{tabular}
\label{tab:renormalization}
\caption{Renormalization constants in the chiral limit at $\beta=1.95$ and $\beta=2.10$ in the $\overline{\rm MS}$-scheme at $\mu=2$~GeV.}
\end{center}
\end{table}
 These are the values that we
use in this work to renormalize the lattice matrix elements. 
  The
numbers in the parenthesis correspond to the statistical error.
Our full results for the renormalization functions of the fermion
field, local and one derivative bilinears along with the systematic
error analysis will appear in a separate publication.

\section{Lattice results}

In this section we present our results on the nucleon electromagnetic form
factors, $G_E(Q^2)$ and $G_M(Q^2)$, and the axial-vector form factors,
$G_A(Q^2)$ and $G_p(Q^2)$. We also show the $n=2$ generalized form factors
for the one-derivative vector operator, $A_{20}(Q^2),\, B_{20}(Q^2)$ and $C_{20}(Q^2)$, 
and the one-derivative axial-vector oprator, $\tilde{A}_{20}(Q^2)$ and $\tilde{B}_{20}(Q^2)$.
The numerical values are given in the Tables in Appendix A. The dependence
of these quantities on the momentum transfer square, $Q^2$, the lattice
spacing, as well as on the pion mass is examined. We also compare
with recent results from other collaborations.

As we already mentioned, most of the results are obtained
for isovector quantities. For the
renormalized nucleon matrix element of the operators with up to one
derivative we thus consider
\beq
\bar u \gamma_{\{\mu} \stackrel{\leftrightarrow}{ D}_{\nu\}} u &-& 
\bar d \gamma_{\{\mu} \stackrel{\leftrightarrow}{ D}_{\nu\}} d\,,\nonumber\\
\bar u \gamma^5\,\gamma_{\{\mu} \stackrel{\leftrightarrow}{ D}_{\nu\}} u &-&
\bar d \gamma^5\,\gamma_{\{\mu} \stackrel{\leftrightarrow}{ D}_{\nu\}} d\,,\nonumber
\label{one-derivative}
\eeq
in the $\overline{\rm MS}$ scheme at a scale $\mu=2$~GeV.
Note that the local vector and axial-vector operators are renormalization
scale independent, thus the conversion to the $\overline{\rm MS}$
scheme is irrelevant.

In order to study  the spin content of the nucleon
we also compute the isoscalar matrix elements of the one-derivative
vector operator, as well as, the isoscalar axial charge assuming, in all cases, that the
disconnected contributions are negligible.

\subsection{Nucleon form factors}

\begin{figure}[h ]
\includegraphics[scale=0.5]{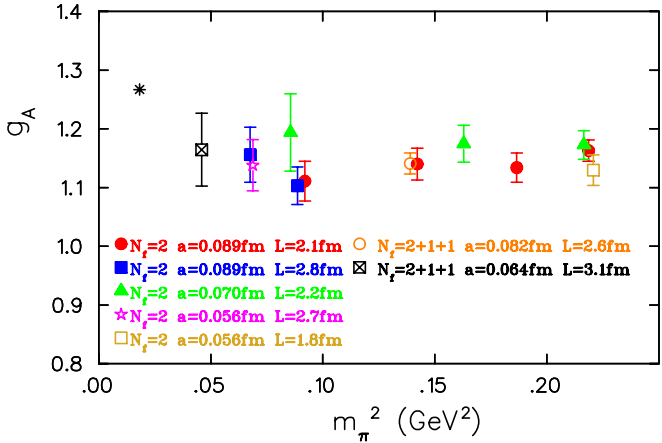}
\caption{Results for the nucleon axial charge with (i) $N_f{=}2$
  twisted mass fermions with $a=0.089$~fm (filled red circles for
  $L=2.1$~fm and filled blue squares for $L=2.8$~fm), $a=0.070$~fm
  (filled green triangles), and $a=0.056$~fm (open star for $L=2.7$~fm
  and open square for $L=1.8$~fm)~\cite{Alexandrou:2010hf} (ii)
  $N_f{=}2{+}1{+}1$ twisted mass fermions with $a=0.082$~fm (open
  circle) and $a=0.064$~fm (square with a cross). The asterisk is the physical
value as given in the PDG~\cite{Beringer:1900zz}. }
\label{gA_mpi_tf}
\end{figure}

In Fig.~\ref{gA_mpi_tf} we present our results for the axial charge
$g_A\equiv G_A(0)$ using 
$N_f{=}2$ and $N_f{=}2{+}1{+}1$ twisted mass fermions.
These are computed at different lattice spacings ranging from $a\sim 0.1$~fm
to $a\sim 0.06$~fm. As can be seen, no sizable cut-off effects
are observed. Lattice data computed using  different volumes are also
consistent down to pion masses of about 300~MeV, where we have
different volumes. In a nutshell, our results do not indicate volume
or cut-off effects larger than our current statistical errors. A
dedicated high statistics analysis using the $N_f{=}2{+}1{+}1$
ensemble at $m_\pi= 373$~MeV has shown that contributions from excited states are negligible for
$g_A$~\cite{Dinter:2011sg,Alexandrou:2011aa}. In  recent studies, the
so called summation method, that sums over the time-slice $t$ where the
current is inserted, is used as an approach that better  suppresses
excited state contributions~\cite{Capitani:2012gj}. Using this method
to analyze lattice results at near physical pion mass it was
demonstrated that, in fact, the value of $g_A$
decreases~\cite{Green:2012ud}. This decrease was
attributed to finite temperature effects~\cite{Green:2012rr}, whereas
for ensembles with large temporal extent the value of $g_A$ was shown to 
increase in accordance with Ref.~\cite{Capitani:2012gj}.
Our main conclusion is that our lattice results are in good agreement with
other lattice computations over the range of pion masses used in this work. 
It is also evident that further investigation is needed to shed light into the behavior of $g_A$ at near physical pion mass. 
\begin{figure}[h ]
\includegraphics[scale=0.5]{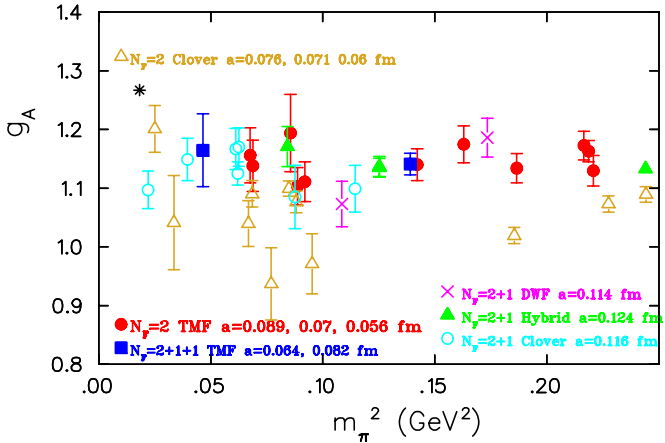}
\caption{The nucleon axial charge for twisted mass fermions, $N_f{=}2$ (filled red circles)
and $N_f{=}2{+}1{+}1$ (filled blue squares), as well as, results using other lattice
actions: Filled (green) triangles correspond to a mixed action with 2+1 flavors
of staggered sea and domain wall valence fermions~\cite{Bratt:2010jn}, crosses to
$N_f{=}2{+}1$ domain wall fermions~\cite{Yamazaki:2009zq}, open triangles
to $N_f{=}2$ clover fermions~\cite{{Horsley:2013ayv}} and  open (cyan)
circles to $N_f{=}2{+}1$ of tree-level clover-improved Wilson fermions
coupled to double HEX-smeared gauge fields~\cite{Green:2012rr}. }
\label{gA_mpi}
\end{figure}
\begin{figure}[h ]
\includegraphics[scale=0.5]{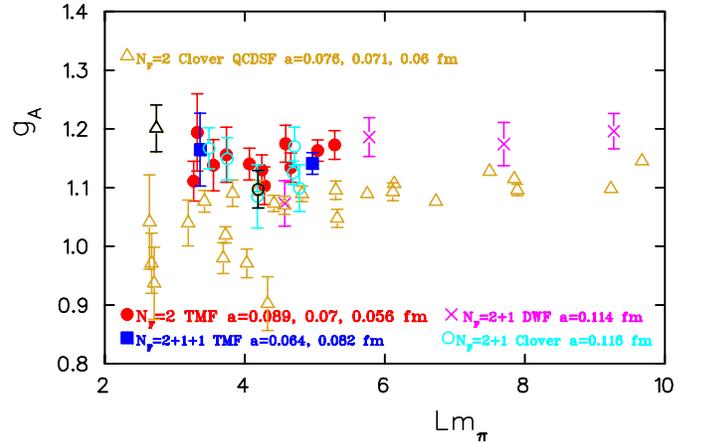}
\caption{The nucleon axial charge for twisted mass fermions ($N_f{=}2$
and $N_f{=}2{+}1{+}1$), as well as results using other lattice
actions versus $Lm_\pi$. Black symbols denote results at almost
physical pion mass obtained using $N_f{=}2$~\cite{Horsley:2013ayv} and
$N_f{=}2{+}1$~\cite{Green:2012rr} clover fermions. The rest of the
notation is the  
same as that in Fig.~\ref{gA_mpi}.}
\label{gA_Lmpi}
\end{figure}

 In Fig.~\ref{gA_mpi}  we compare our results to other recent lattice QCD data
obtained with different actions. We show results obtained using
 domain wall fermions (DWF)~\cite{Yamazaki:2009zq},
clover fermions~\cite{Horsley:2013ayv}, a mixed action with 2+1 flavors of asqtad-improved staggered
sea and domain wall valence fermions~\cite{Bratt:2010jn} referred to as hybrid, and
$N_f{=}2{+}1$ of tree-level clover-improved Wilson fermions coupled to
double HEX-smeared gauge fields~\cite{Green:2012ud, Green:2012rr}. We
observe that all these lattice results are compatible. This agreement 
corroborates the fact that cut-off effects are negligible since these lattice
data are obtained with different discretized actions without being
extrapolated to the continuum limit. The recent result of
Ref.~\cite{Green:2012rr}  at almost physical pion mass shows about
  $10 \%$ deviation
from the physical value of $g_A^{exp}=1.267$~\cite{Beringer:1900zz}. This
is a well-known puzzle  and various directions
have been explored to identify the source of the
discrepancy~\cite{Capitani:2010sg, Green:2011fg,
  Alexandrou:2011aa,Dinter:2011sg}. In Fig.~\ref{gA_mpi} we also
include the recent results obtained using $N_f{=}2$ clover fermions at
three lattice spacings $a=0.076$~fm, 0.071~fm and
0.060~fm~\cite{Horsley:2013ayv}. They include a result at almost
physical pion mass, which is clearly higher than the corresponding one
obtained in Ref.~\cite{Green:2012rr}. As already remarked, the latter  was  shown  to
even decrease if one uses the summation method~\cite{Green:2012ud}. 
In Ref.~\cite{Horsley:2013ayv} it is argued that volume corrections are sizable
and increase the value of $g_A$.
We note that all lattice data shown in Fig.~\ref{gA_mpi} are not volume corrected. In order
to assess, which of these results would suffer from large volume corrections
we show in Fig.~\ref{gA_Lmpi} $g_A$ as a function of $Lm_\pi$. The data points
at almost physical pion mass are shown with the black symbols. The
result from Ref.~\cite{Green:2012rr} at $Lm_\pi=4.2$ is lower than the
one from Ref.~\cite{Horsley:2013ayv} at $Lm_\pi=2.74$. Thus volume
effects alone may not account for the whole discrepancy and
therefore, there is still an open issue in the evaluation of $g_A$. 

\begin{figure}[h]
{\includegraphics[scale=0.3]{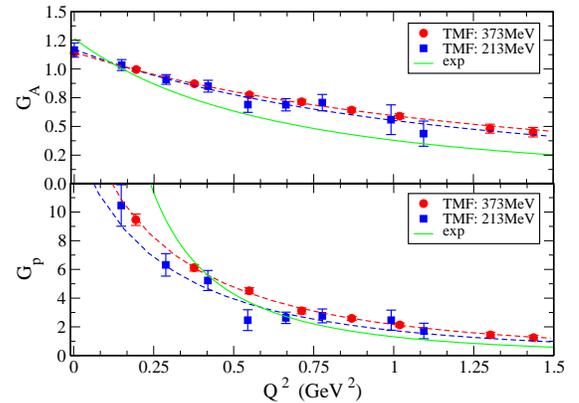}}
\caption{Comparison  of the $N_f{=}2{+}1{+}1$ twisted mass data
  on $G_A(Q^2)$ (upper) and $G_p(Q^2)$ (lower) for the two different pion masses considered. 
Filled blue squares correspond to $\beta=2.10$ and
  $m_\pi= 213$~MeV, while filled red circles correspond to $\beta=1.95$ and
$m_\pi= 373$~MeV. The dashed lines are the dipole fits on the
  lattice data, while the solid green line is the dipole fit 
of experimental data for $G_A(Q^2)$~\cite{Bernard:2001rs}  in combination with pion-pole
dominance for $G_p(Q^2)$.}\label{fig:GAGp_mpi}
\end{figure}
Next, we study the dependence of the axial form factors on the momentum
transfer, $Q^2$. 
In Fig.~\ref{fig:GAGp_mpi} we compare our $N_f{=}2{+}1{+}1$ 
results for $G_A(Q^2)$ and $G_p(Q^2)$ as the pion mass decreases from
373~MeV to 213~MeV. As can be seen, the dependence on the pion mass is
very weak for $G_A(Q^2)$ whereas for $G_p(Q^2)$ a stronger dependence
is observed in particular at low $Q^2$. This is not surprising since
$G_p(Q^2)$ is expected to have a pion-pole dependence that dominates
its $Q^2$-dependence as $Q^2\to 0$. The solid line is the result of a
dipole fit to the experimental electroproduction data for
$G_A(Q^2)$. Assuming pion-pole dominance we can deduce from the fit to
the experimental data on $G_A(Q^2)$ the expected behavior for
$G_p(Q^2)$, shown in Fig.~\ref{fig:GAGp_mpi}. As can be seen,  both
quantities have a smaller slope with respect to $Q^2$ than what is
extracted from experiment. Such a behavior is common to all the
nucleon form factors and it remains to be further investigated if
reducing even more the pion mass will resolve this discrepancy. The
$Q^2$-dependence of the lattice QCD data  for $G_A(Q^2)$ can be well
parameterized by dipole Ansatz of the form
\be
G_A(Q^2) = \frac{g_A}{\left(1+Q^2/m^2_A\right)^2}\>,
\ee
as it was done for the experimental results.
Likewise, assuming  pion-pole
dominance we fit $G_p(Q^2)$ to the form
\be
G_p(Q^2) = \frac{G_A(Q^2)\,G_p(0)}{\left(Q^2 +m^2_p \right)}\,.
\ee
In both fits we take into account lattice data with $Q^2$
 up to a maximum value of $(1.5)^2$~GeV$^2$. The values of the parameter $m_A$ 
extracted from the fit for the two ensembles are 
\beq
\beta & = &1.95:\,\,m_A= 1.60(5)~{\rm GeV} \nonumber \\
\beta& = &2.10:\,\,m_A= 1.48(12)~{\rm GeV}\, . \nonumber 
\eeq
These are higher than the experimental value of $m^{exp}_A$ = 1.069~GeV~\cite{Bernard:2001rs} 
extracted from the best dipole parameterization to the
electroproduction data. This deviation between lattice and
experimental data reflects the smaller slope in the lattice QCD data.
Another observation is that the fits for $G_p(Q^2)$ are strongly dependent on the lowest values
of $Q^2$ taken in the fit due to the strong $Q^2$-dependence of $G_p(Q^2)$ at low $Q^2$.

\begin{figure}[h]
{\includegraphics[scale=0.3]{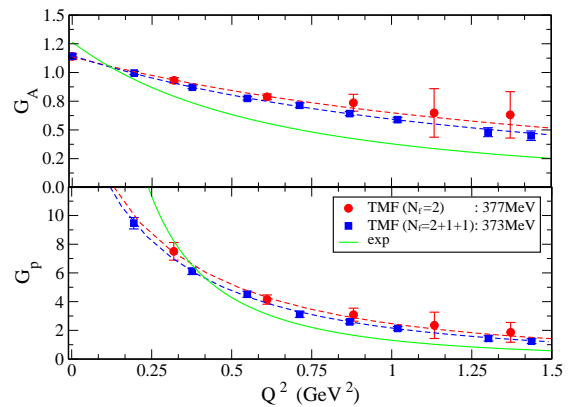}}
\caption{The $Q^2$-dependence of the form factors $G_A$ and $G_p$ for
 i) $N_f{=}2$ at $m_\pi=377$~MeV, $a=0.089$~fm (filled red circles);
 ii) $N_f{=}2{+}1{+}1$ at $m_\pi=373$~MeV, $a=0.082$~fm. The solid
 line in the upper plot shows the resulting dipole fit to  the
 experimental data on $G_A(Q^2)$~\cite{Bernard:2001rs}. Assuming a
 pion-pole dependence for $G_p(Q^2)$ and using the fit on $G_A(Q^2)$
 shown in the upper panel produces the solid line shown in the lower
 panel for $G_p(Q^2)$.}
\label{GAGp_Q2_b1.95}
\end{figure}

\begin{figure}[h]
{\includegraphics[scale=0.3]{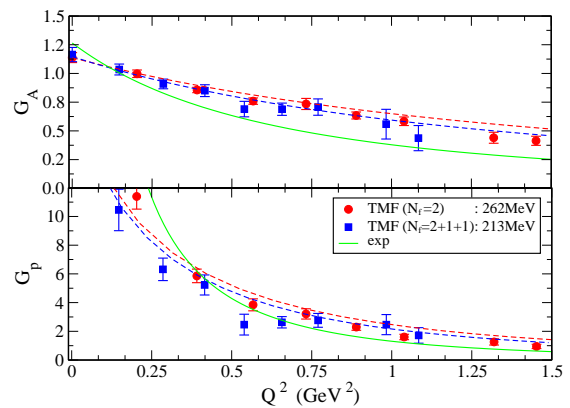}}
\caption{The $Q^2$-dependence of the form factors $G_A$ (upper) and
  $G_p$ (lower) for $N_f{=}2{+}1{+}1$ twisted mass fermions at
  $m_\pi=213$~MeV, $a=0.064$~fm (filled blue squares) and $N_f{=}2$
  twisted mass fermions at $m_\pi=262$~MeV and $a=0.056$~fm (filled
  red circles).  The rest of the notation is the same as that in
  Fig.~\ref{GAGp_Q2_b1.95}. }
\label{GAGp_Q2_b2.1}
\end{figure}

In Figs.~\ref{GAGp_Q2_b1.95} and \ref{GAGp_Q2_b2.1} we
compare results using the  two
 $N_f{=}2{+}1{+}1$ ensembles with those obtained with $N_f{=}2$ ensembles at similar pion masses. We do not observe
large deviations between $N_f{=}2$ and  $N_f{=}2{+}1{+}1$ result
showing that strange and charm quark effects are small, as expected.

\begin{figure}[h]
{\includegraphics[scale=0.5]{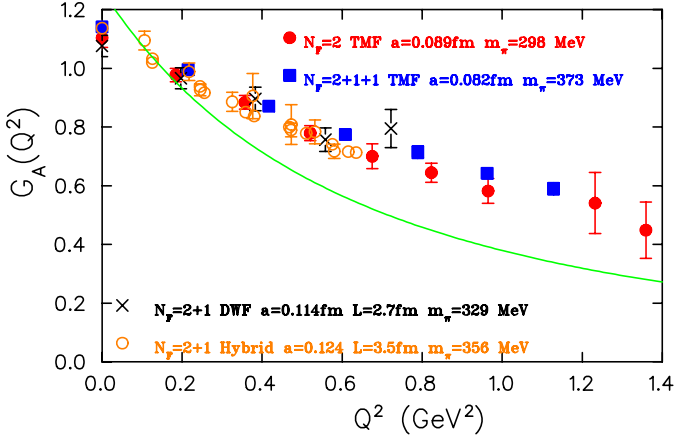}}
\caption{ $Q^2$-dependence of $G_A(Q^2)$ for $N_f{=}2{+}1{+}1$  at $m_\pi=373$~MeV (filled blue squares) and the
  $N_f{=}2$~\cite{Alexandrou:2010hf} at $m_\pi=298$~MeV  (filled red
  circles) twisted mass data on a lattice with spatial length
  $L=2.8$~fm and similar lattice spacing.  We also show results with
  $N_f{=}2{+}1$ DWF at $m_\pi=329$~MeV, $L=2.7$~fm
  (crosses)~\cite{Yamazaki:2009zq} and with a hybrid action with
  $N_f{=}2{+}1$ staggered sea and DWF 
at $m_\pi=356$~MeV and $L=3.5$~fm (open orange circles)~\cite{Bratt:2010jn}.}\label{fig:GA_compare} 
\end{figure}

\begin{figure}[h]
{\includegraphics[scale=0.5]{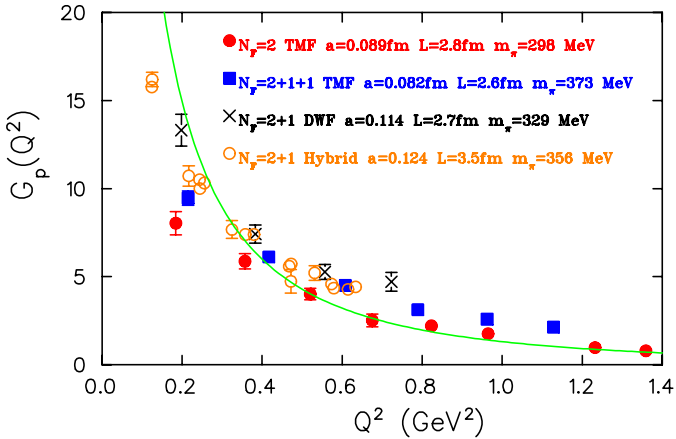}}
\caption{The $Q^2$-dependence of $G_p(Q^2)$. The notation is the same as that in Fig.~\ref{fig:GA_compare}.}\label{fig:Gp_compare} 
\end{figure}

It is interesting to compare our TMF results to those obtained using
different fermion discretization schemes. We collect recent lattice
QCD results in Figs.~\ref{fig:GA_compare} and \ref{fig:Gp_compare} at
similar pion masses. As can be seen, in the case of $G_A(Q^2)$ there
is agreement of our results with those obtained using DWF and the hybrid
approach. For $G_p(Q^2)$ hybrid
results obtained on a larger volume are higher at small
$Q^2$-values. This is an indication that volume effects are larger for
quantities like $G_p(Q^2)$ for which pion cloud effects are expected
to be particularly large at small $Q^2$.

\begin{figure}[h]
{\includegraphics[scale=0.3]{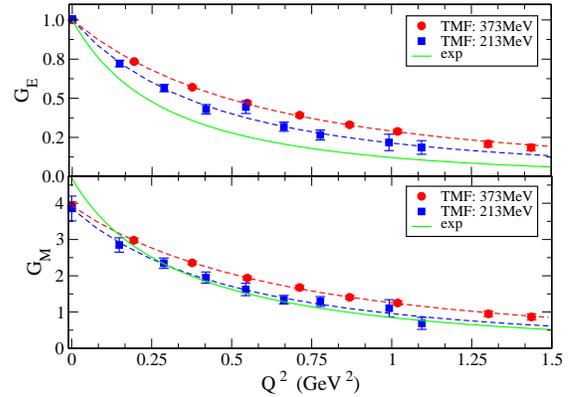}}
\caption{Comparison  of the $N_f{=}2{+}1{+}1$ twisted mass data
  on $G_E(Q^2)$ (upper) and $G_M(Q^2)$ (lower) for the two different
  pion masses considered. The solid lines are Kelly's parameterization
  of the experimental data~\cite{Kelly:2004hm}, whereas the dashed lines are dipole fits
  to the lattice QCD data. 
}
\label{fig:GEGM_mpi}
\end{figure}

We next discuss the results obtained for the isovector electromagnetic
form factors, $G_E(Q^2)$ and $G_M(Q^2)$. 
In Fig.~\ref{fig:GEGM_mpi} we compare our $N_f{=}2{+}1{+}1$ results as the
pion mass decreases from 373~MeV to 213~MeV.  As can be seen,  the
values for both quantities  decrease towards the experimental values
shown by the solid line, which is J. Kelly's parameterization to the
experimental data~\cite{Kelly:2004hm}. In particular, for $G_M(Q^2)$
lattice results at $m_\pi=213$~MeV become consistent with the
experimental results. In order to extract the value of $G_M(0)$, we
need to extrapolate lattice results at finite $Q^2$. We parameterized
both form factors by a dipole form
\beq
G_E(Q^2){=}\frac{1}{(1{+}Q^2/m_E^2)^2}\,, \nonumber \\
 G_M(Q^2){=}\frac{G_M(0)}{(1{+}Q^2/m_M^2)^2}.
\label{dipole}
\eeq
The values of $G_M(0)$ extracted are shown in Fig.~\ref{fig:GEGM_mpi},
as well as, the resulting fits  with the dashed lines.  The overall trend of the lattice QCD data clearly shows that as the pion mass
decreases they approach the experimental values. However, even at
$m_\pi=213$~MeV the value of $G_M(0)$, which determines the isovector
anomalous magnetic moment, is still underestimated. In
Table~\ref{Table:GE GM fits} we tabulate the resulting fit parameters
$m_E$, $G_M(0)$ and $m_M$ for the two $N_f{=}2{+}1{+}1$ ensembles
extracted from the dipole fits of Eqs.~(\ref{dipole}).
  \begin{table}[h]
\begin{center}
\begin{tabular}{c|c|c|c}
\hline\hline
$\beta$ & $m_E$ (GeV)& $G_M(0)$ & $m_M$ (GeV)  
\\\hline
1.95 &  1.17(32)  & 3.93(12)   & 1.30(08)     \\
2.10 &  0.86(07)  & 3.86(34)   & 0.99(15)     \\
\hline
\end{tabular}
\caption{Results on the nucleon electric and magnetic mass extracted
  by fitting to the dipole form of Eq.~(\ref{dipole}).}
\label{Table:GE GM fits}
\end{center}
\end{table}

\begin{figure}[h]
{\includegraphics[scale=0.3]{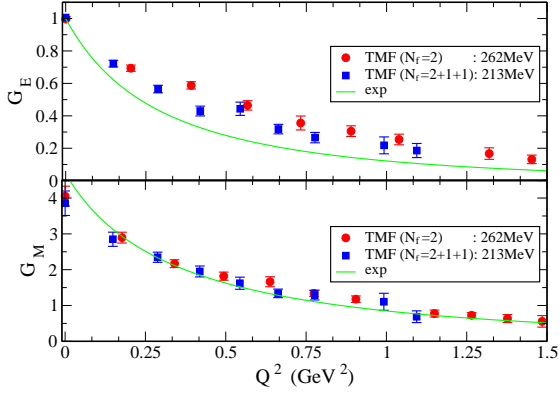}}
\caption{The $Q^2$-dependence of $G_E(Q^2)$ (upper) and $G_M (Q^2)$ (lower) for $N_f{=}2{+}1{+}1$ TMF
at $m_\pi=213$~MeV (filled blue squares) and $N_f{=}2$ TMF at $m_\pi=262$~MeV~(filled red circles). }
\label{GEGM_Q2_b2.1}
\end{figure}

In Fig.~\ref{GEGM_Q2_b2.1}
we show the $Q^2$ dependence of $G_E(Q^2)$ and $G_M(Q^2)$ 
at $\beta=2.10$ and $m_\pi=213$~MeV comparing it to the smallest
available pion mass of $262$~MeV  obtained using $N_f{=}2$ ensembles.
Once again we do not observe any sizable effects due to the strange and charm quarks in the sea.

\begin{figure}[h]
{\includegraphics[scale=0.5]{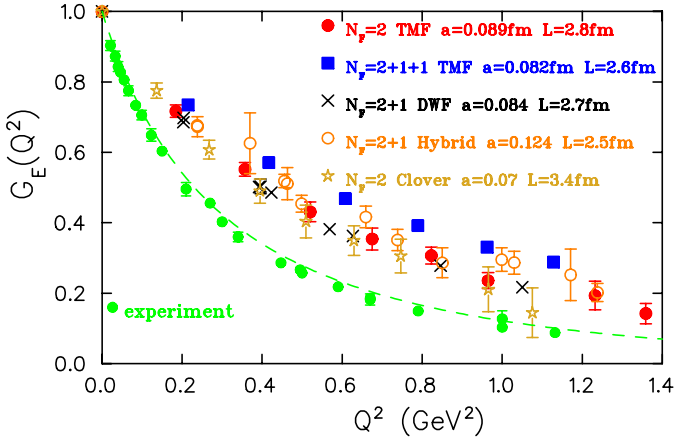}}
\caption{The $Q^2$-dependence of $G_E(Q^2)$.
 We show results for $N_f{=}2{+}1{+}1$ at $m_\pi=373$~MeV
 (filled blue squares) and $N_f{=}2$~\cite{Alexandrou:2011nr} at
 $m_\pi=298$~MeV (filled red circles) TMF  data on a lattice
 with spatial length  $L=2.8$~fm and similar lattice spacing.  We also
 show results with $N_f{=}2{+}1$ DWF at $m_\pi=297$~MeV, $L=2.7$~fm
 (crosses)~\cite{Syritsyn:2009mx}, with a hybrid action with $N_f{=}2{+}1$
 staggered sea and DWF at $m_\pi=293$~MeV and $L=2.5$~fm (open orange
 circles)~\cite{Bratt:2010jn},  and $N_f{=}2$ clover at $m_\pi=290$~MeV
 and $L=3.4$~fm~(asterisks)~\cite{Capitani:2010sg}. The solid line is Kelly's parameterization
  of the experimental data~\cite{Kelly:2004hm} from a number of experiments
as given in Ref.~\cite{Kelly:2004hm}.}
\label{fig:GE_compare}
\end{figure}

\begin{figure}[h]
{\includegraphics[scale=0.5]{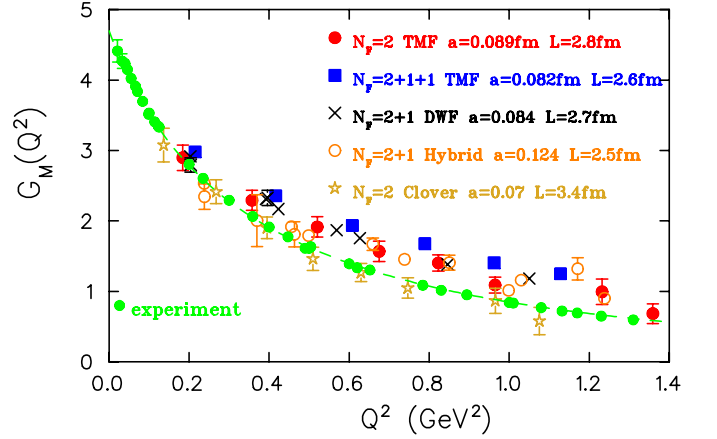}}
\caption{The $Q^2$-dependence of $G_M(Q^2)$.
The notation is the same as that in Fig.~\ref{fig:GE_compare}.
}
\label{fig:GM_compare}
\end{figure}

It is useful to  compare TMF results to those
obtained within different fermion discretization schemes.  
In particular, we compare in Figs.~\ref{fig:GE_compare} and \ref{fig:GM_compare}with  
results obtained using $N_f{=}2{+}1$ DWF~\cite{Syritsyn:2009mx},
$N_f{=}2$ Wilson improved clover fermions~\cite{Capitani:2010sg} and
using the  hybrid action~\cite{Bratt:2010jn} for a pion
mass of about 300~MeV. We see a nice agreement among all lattice
results for $G_E(Q^2)$, confirming that cut-off effects are small for these
actions. In the case of $G_M(Q^2)$ there is also an overall agreement
except in the case of the $N_f{=}2$ clover results. These results are somewhat lower
and are more in agreement with our results at $m_\pi=213$~MeV. The
reason for this is unclear and might be due to limited statistics as
these data carry the largest errors.  


\begin{figure}\vspace*{0.3cm}  
\includegraphics[width=\linewidth]{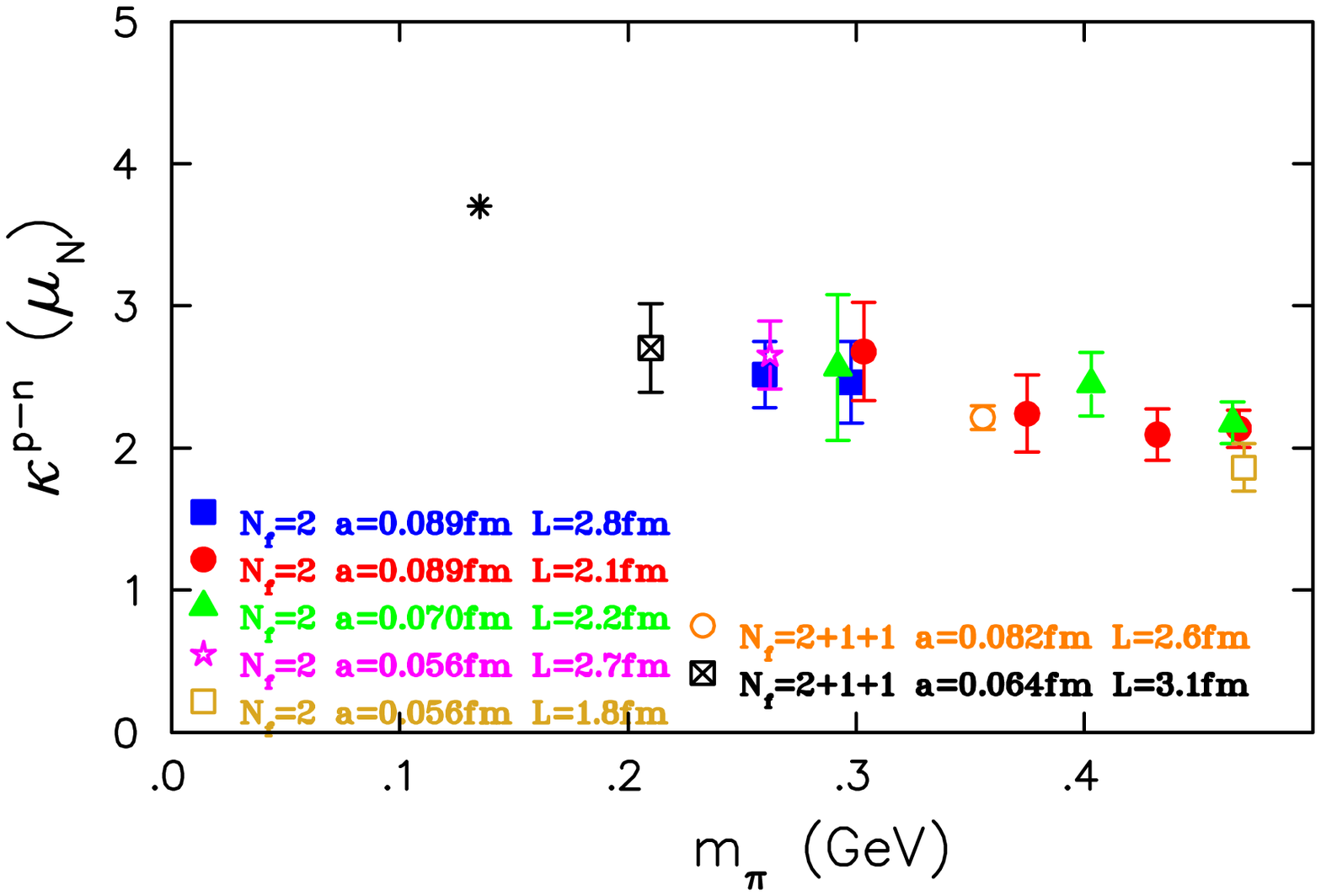}\\  
\includegraphics[width=\linewidth]{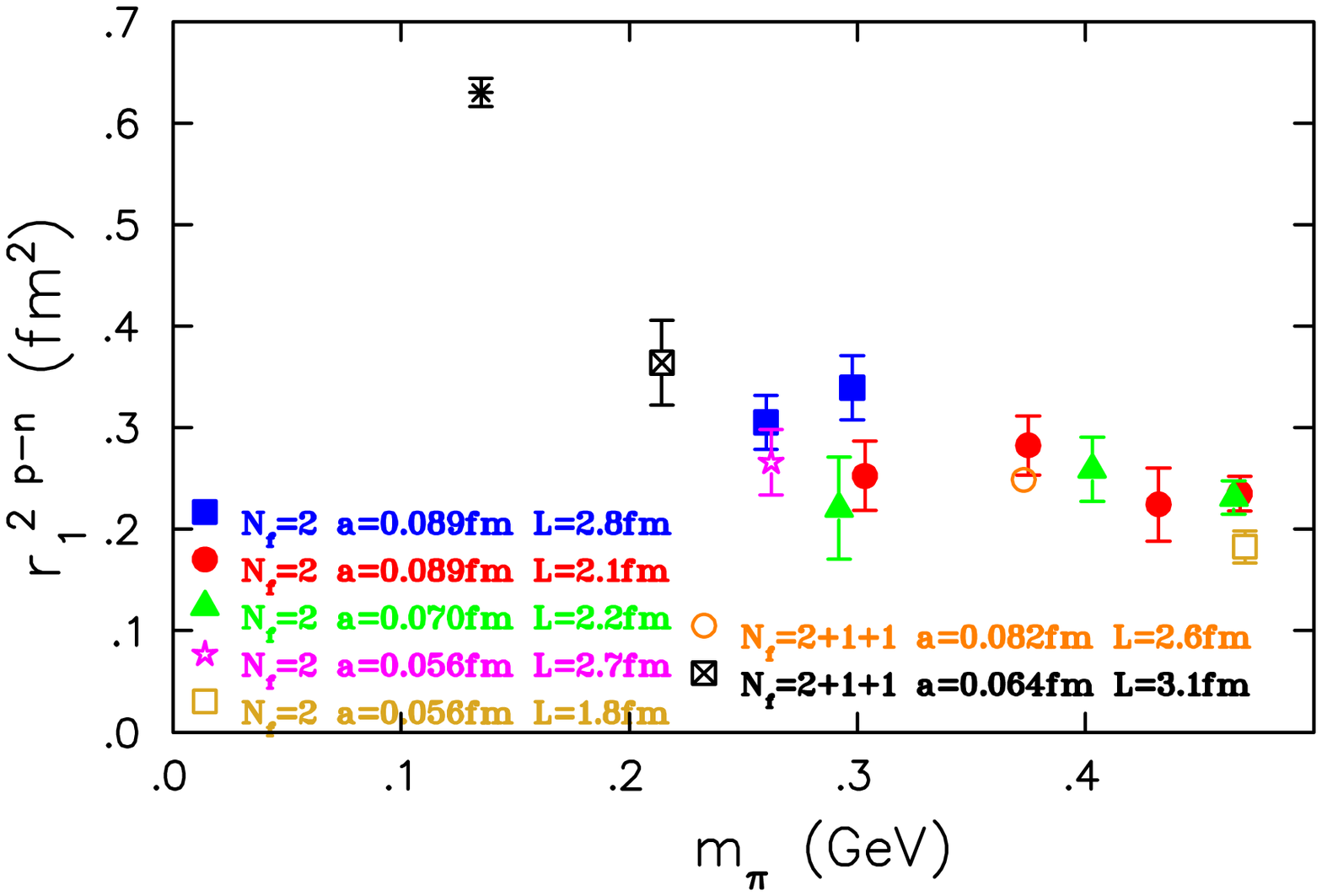}\\  
\includegraphics[width=\linewidth]{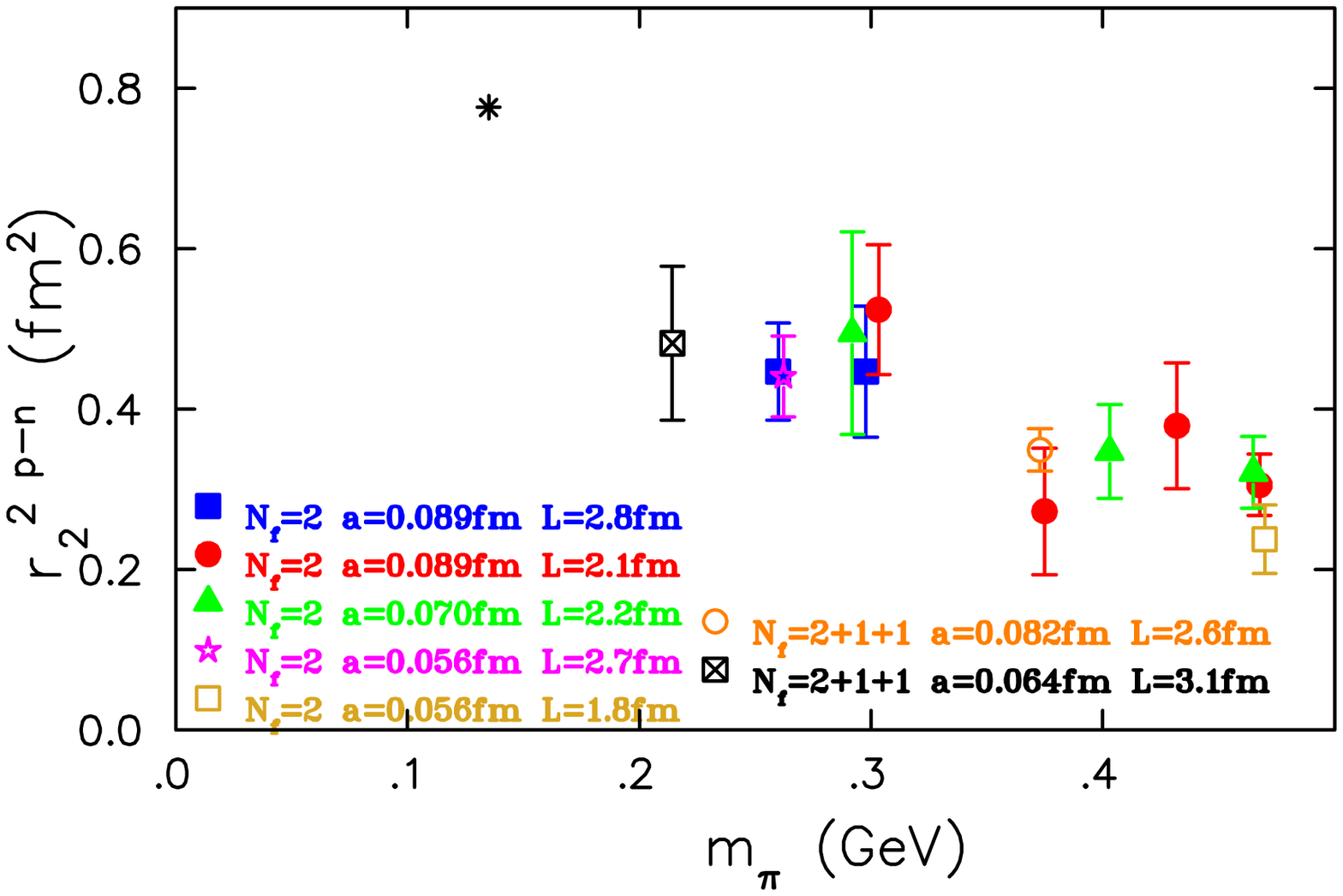}  
\caption{Twisted mass fermion results with $N_f{=}2$~\cite{Alexandrou:2010hf}   
and with $N_f{=}2{+}1{+}1$, for the isovector  anomalous magnetic
moment, $\kappa^{p-n}$ in Bohr magnetons (upper), Dirac r.m.s. radius (middle) and Pauli r.m.s. radius
(lower) panel. The notation is the same as that in
Fig.~\ref{gA_mpi_tf}.} 
\label{fig:kappav r1 r2}  
\end{figure}

Having fitted the electromagnetic form factors we can extract the isovector anomalous magnetic moment and root mean square (r.m.s.) radii.
The anomalous magnetic moment is given by the Pauli
form factor $F_2(0)$ and 
the slope of $F_1$ at $Q^2=0$ determines the transverse size of the hadron,
$\langle r_\perp^2\rangle = -4 dF_1/dQ^2|_{Q^2=0}$. In
the non-relativistic limit  the r.m.s. radius is related to
the slope of the form factor at zero momentum transfer. Therefore 
the r.m.s. radii can
 be obtained  from the values of the dipole masses by using
\be
\langle r_i^2 \rangle=-\frac{6}{F_i(Q^2)}\frac{dF_i(Q^2)}{dQ^2}|_{Q^2=0}=\frac{12}{m_i^2}, \hspace*{0.5cm} i=1,2 \quad.
\ee
The electric and magnetic radii are 
 given by $\langle r_{E,M}^2\rangle=12/m_{E,M}^2$ and can be directly evaluated from
the values of the parameters listed
in Table~\ref{Table:GE GM fits}.
In Fig.~\ref{fig:kappav r1 r2} we present our results on the anomalous magnetic moment, Dirac and  Pauli  r.m.s. radii.
As can be seen, the new results at $m_\pi=213$~MeV, although they are still lower
than the experimental value, show an increase towards that value.
In Ref.~\cite{Green:2012ud} an analysis  of the results using the 
summation method at $m_\pi=147$~MeV with $N_f{=}2{+}1$ clover fermions
was carried out. It was shown that the value of these three quantities
increases to bring agreement with the experimental value. This is an
encouraging result that needs to be  confirmed.

\subsection{Nucleon generalized form factors with one derivative operators}

In this section we present results on the nucleon matrix elements of
the isovector one-derivative operators defined in
Eq.~(\ref{one-derivative}). The full body of our results are collected
in Tables~\ref{tab:results_ffs} and \ref{tab:results_Gffs} in Appendix
A. Like $g_A$, $A_{20}(Q^2{=}0)$ and $\tilde A_{20}(Q^2{=}0)$ can be
extracted directly from the corresponding matrix element at
$Q^2=0$. On the other hand, $B_{20}(Q^2{=}0)$, $C_{20}(Q^2{=}0)$ and
$\tilde B_{20}(Q^2{=}0)$, like $G_M$ and $G_p$, can not be extracted
at $Q^2=0$. Therefore one needs to extrapolate lattice data at
$Q^2{\neq}0$ by performing a fit.

\begin{figure}
{\includegraphics[scale=0.5]{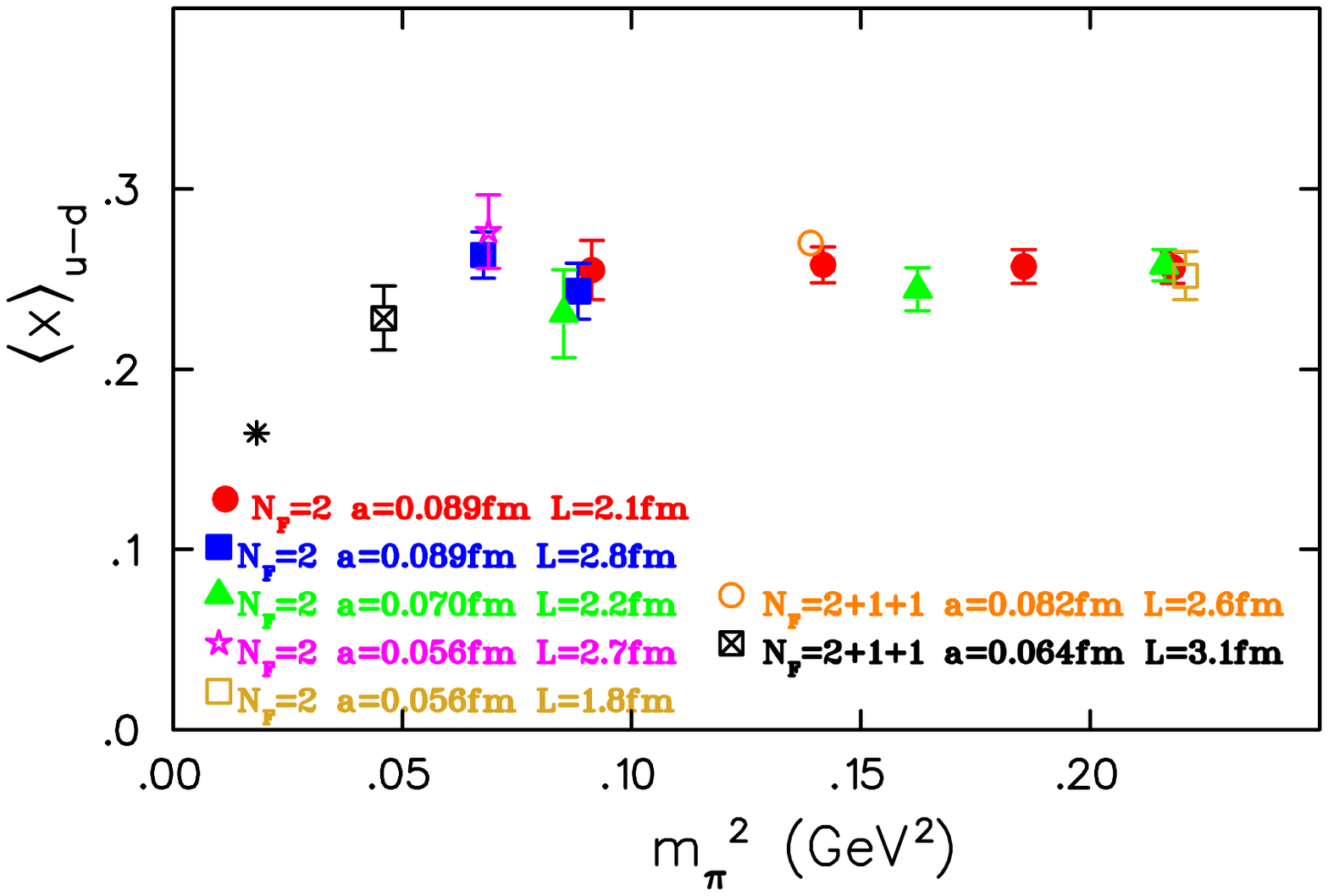}}\\[2ex]
{\includegraphics[scale=0.5]{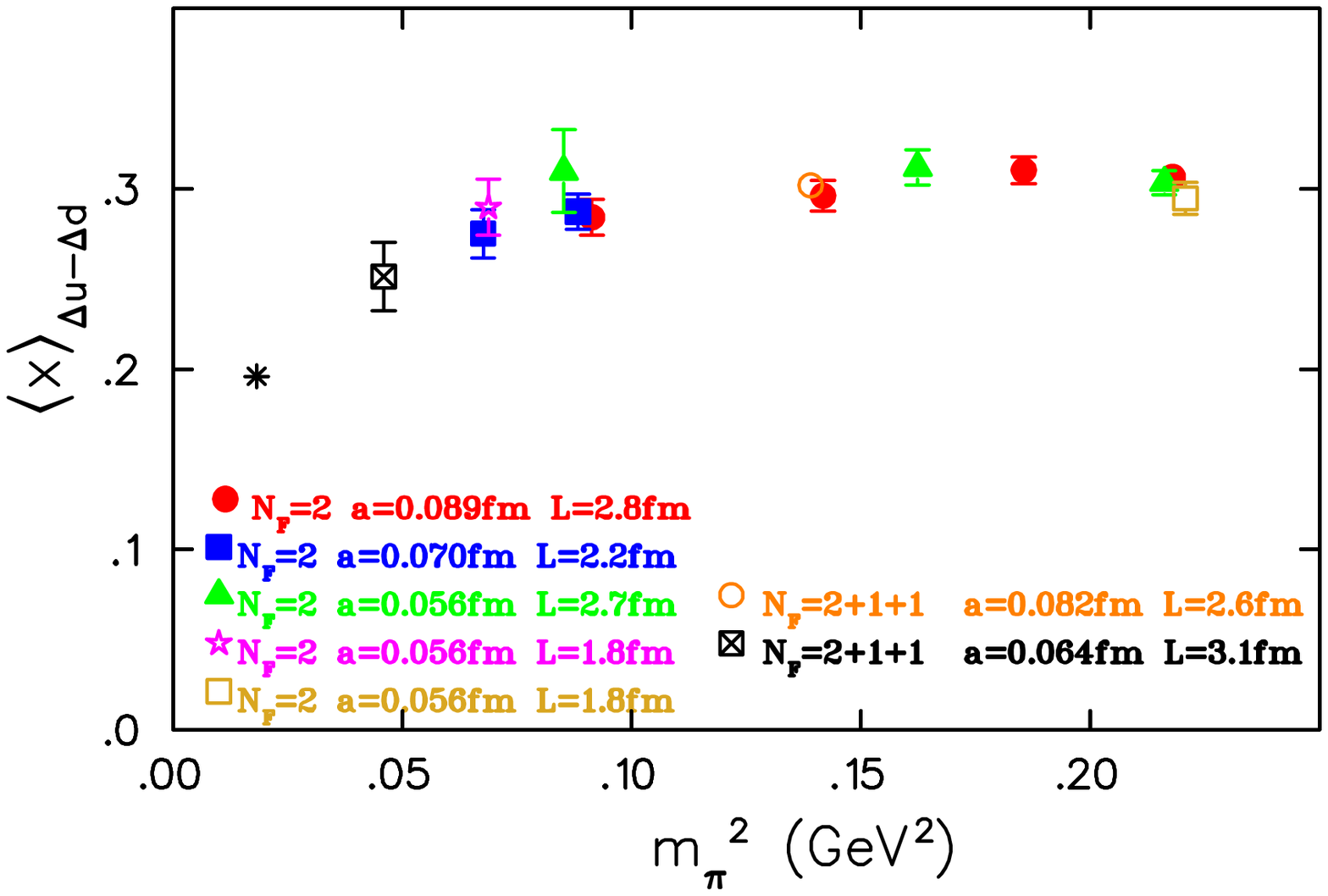}}
\caption{Results for $\langle x \rangle_{u-d}$ (upper) and  $\langle x \rangle_{\Delta u-\Delta d}$ (lower)
using $N_f{=}2$ and $N_f{=}2{+}1{+}1$ twisted mass fermions as a function of the pion mass.
 We show results for (i) $N_f{=}2$ twisted mass fermions with
 $a=0.089$~fm (filled red circles for $L=2.1$~fm and filled blue
 squares for $L=2.8$~fm), $a=0.070$~fm  (filled green triangles), and
 $a=0.056$~fm (open stars for $L=2.7$~fm and open square for
 $L=1.8$~fm); (ii) $N_f{=}2{+}1{+}1$ twisted mass fermions with
 $a=0.0820$~fm (open circle) and $a=0.0657$~fm (square with a
 cross). The physical point, shown by the asterisk, is from
 Ref.~\cite{Alekhin:2009ni} for the unpolarized and from
 Ref.~\cite{Airapetian:2009ac,Blumlein:2010rn} for the polarized first
 moment. } 
\label{fig:A20 ETMC}
\end{figure}

In Fig.~\ref{fig:A20 ETMC}  we compare our lattice data of the
unpolarized and polarized isovector moments obtained for
$N_f{=}2$~\cite{Alexandrou:2011nr} TMF for different lattice spacings and
volumes to the $N_f{=}2{+}1{+}1$ TMF results of this work.
  As can be seen, there are no detectable cut-off
effects for the lattice spacings considered here, nor  volume
dependence at least for pion masses up to about 300~MeV where
different volumes were analyzed.  Also, there is consistency among
results obtained using $N_f{=}2$ and $N_f{=}2{+}1{+}1$ gauge
configurations indicating that strange and charm quark effects are
small.
We would like to point out  that the renormalization constant for the vector one-derivative operator 
is larger by about 2\% than the one used in Ref.~\cite{Dinter:2011sg} since
in converting to $\overline{\rm MS}$ we used the 2-loop conversion factor instead of the  3-loop result, thus
increasing the value of $\langle x \rangle_{u-d}$. 
 As in the case of the nucleon axial charge, a number of studies were
undertaken to examine the role of excited states in the extraction of
$\langle x \rangle_{u-d}$. A high statistics analysis carried out with
twisted mass fermions at $m_\pi= 373$~MeV has shown that excited
state contamination accounted for a decrease of about 10\% in the
value of $\langle x \rangle_{u-d}$ as compared to the value extracted using sink-source
separation of about 1~fm~\cite{Dinter:2011sg,Alexandrou:2011aa}.
The most noticeable behavior regarding these TMF results is that the values
obtained  at $m_{\pi}=213$~MeV for both $\langle x\rangle_{u-d}$ and
 $\langle x\rangle_{\Delta u-\Delta d}$ approach the physical value.
We would like to remark that the phenomenological value of $\langle
x\rangle_{u-d}$ extracted from different
analysis~\cite{Blumlein:2004ip,Blumlein:2006be,Alekhin:2006zm,Alekhin:2009ni,JimenezDelgado:2008hf,Martin:2009bu}
shows a spread, which, however, is significantly smaller than the
discrepancy as compared to the deviation shown by lattice data for pion masses
higher than the physical point. The same applies for $\langle
x\rangle_{\Delta u-\Delta d}$~\cite{Airapetian:2009ac,Blumlein:2010rn}.
 
\begin{figure}
{\includegraphics[scale=0.5]{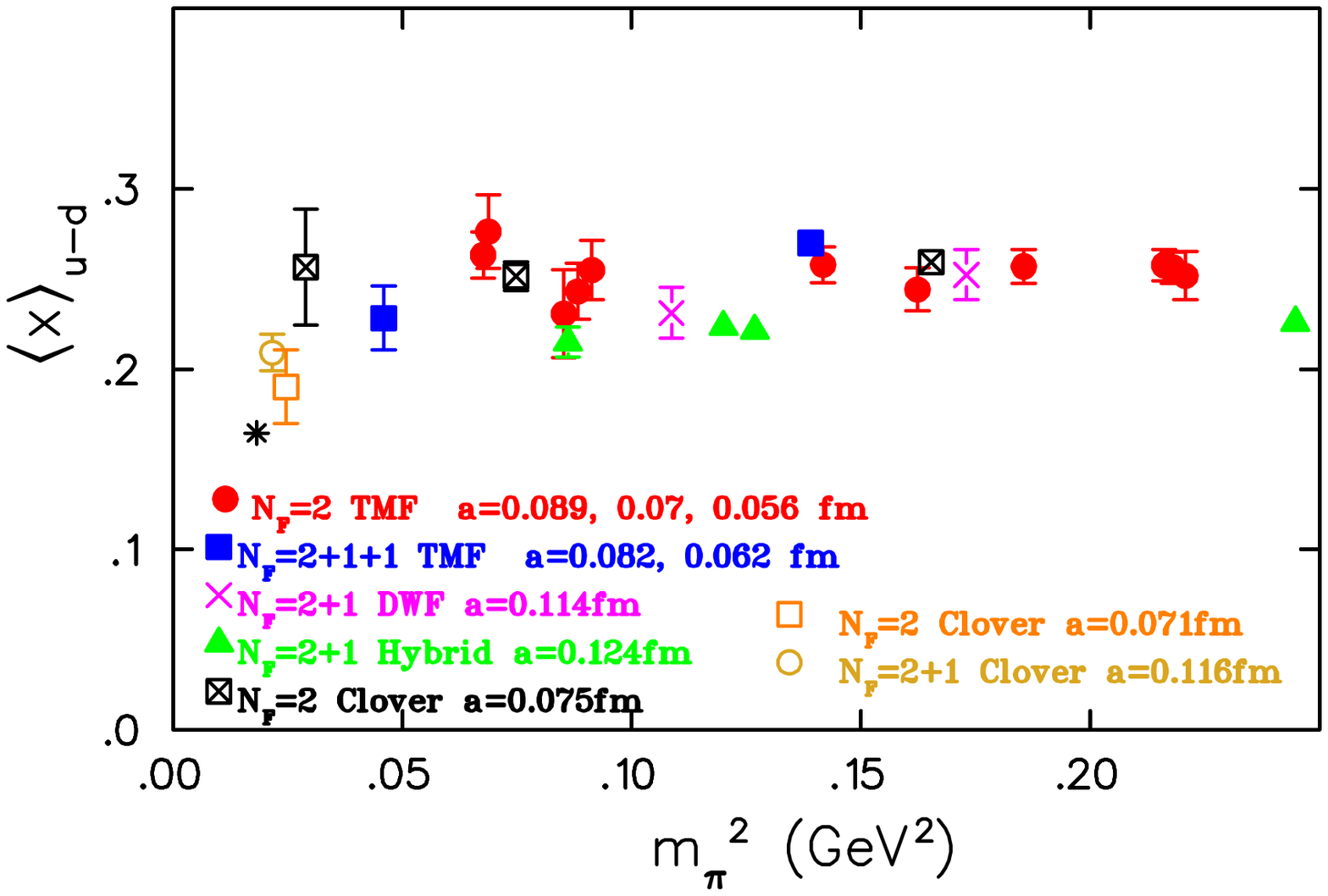}}\\[2ex]
{\includegraphics[scale=0.5]{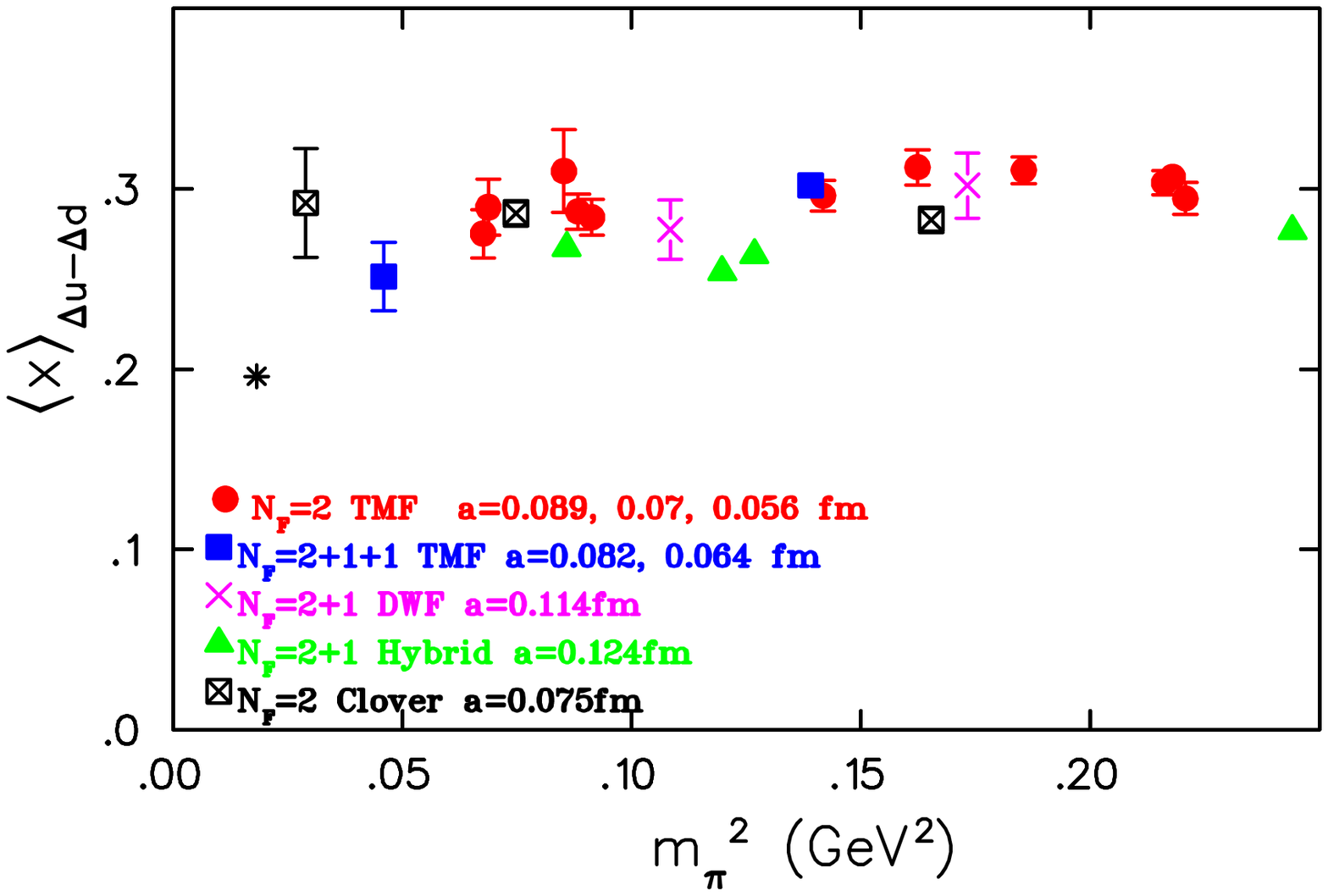}}
\caption{Results for $\langle x \rangle_{u-d}$ (upper) and  $\langle x
  \rangle_{\Delta u-\Delta d}$ (lower) obtained in this work are shown
  with the red filled circles for $N_f{=}2$ and with the blue filled
  squares for $N_f{=}2{+}1{+}1$. We compare with (i) $N_f{=}2{+}1$ DWF
  for $a=0.114$~fm~\cite{Aoki:2010xg}; (ii) $N_f{=}2{+}1$ using DWF
  for the valence quarks on staggered sea~\cite{Bratt:2010jn} with $a=0.124$~fm; 
(iii) $N_f{=}2$ clover with $a=0.075$~fm~\cite{Pleiter:2011gw}. For
 $\langle x \rangle_{u-d}$ we also show recent results using 
 $N_f{=}2$ clover with $a=0.071$~fm~\cite{Bali:2012av} and
 $N_f{=}2{+}1$ of tree-level clover-improved Wilson fermions coupled
 to double HEX-smeared gauge fields with
 $a=0.116$~fm~\cite{Green:2012ud}.}
\label{fig:A20A20tilde_all}
\end{figure}

\begin{figure}\vspace*{0.5cm}
{\includegraphics[scale=0.35]{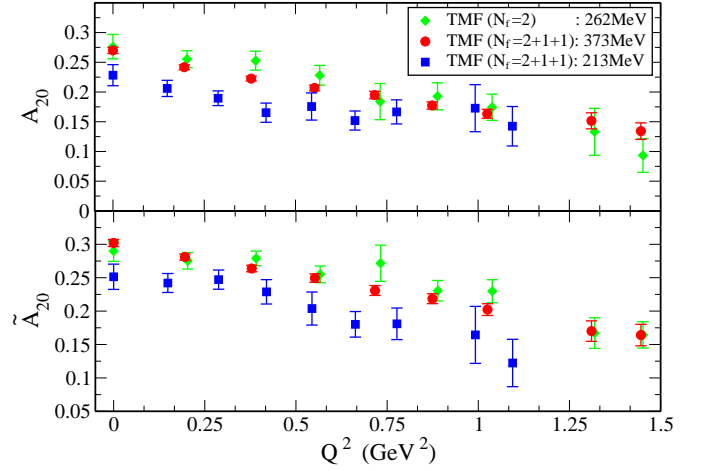}}
 \caption{The $Q^2$-dependence of  $A_{20}(Q^2)$ (upper) and $\tilde{A}_{20}(Q^2)$ (lower) for $N_f{=}2$ with 
$a=0.056$~fm and $m_\pi=262$~MeV (filled green diamonds), and $N_f{=}2{+}1{+}1$ with
i) $a=0.064$~fm and $m_\pi=213$~MeV (filled blue squares); ii) $a=0.082$~fm and $m_\pi=373$~MeV (filled red circles). }
\label{fig:A20A20tilde_tmf}
\end{figure}

\begin{figure}
{\includegraphics[scale=0.35]{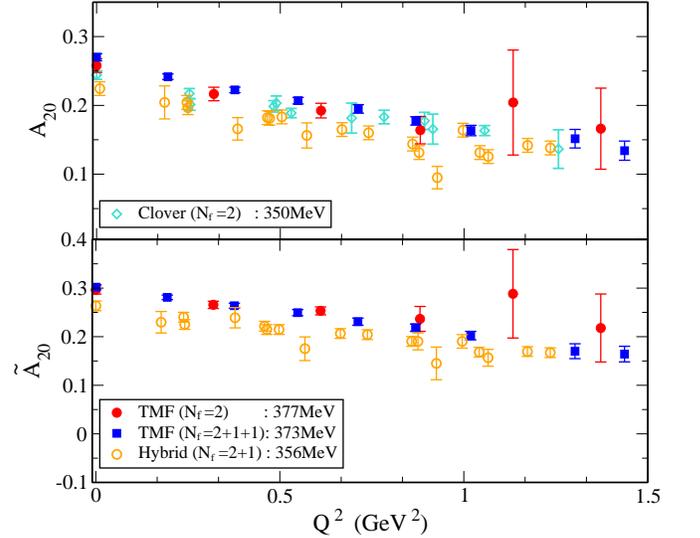}}
\caption{The $Q^2$-dependence of  $A_{20}(Q^2)$ (upper) and $\tilde{A}_{20}(Q^2)$
  (lower) shown for i) $N_f{=}2$ twisted mass fermions for
  $a=0.089$~fm, $m_\pi=377$~MeV (filled red
  circles)~\cite{Alexandrou:2011nr}; ii) $N_f{=}2{+}1{+}1$ twisted
  mass fermions (this work) for $a=0.082$~fm and $m_\pi=373$~MeV;
  iii) $N_f{=}2$ clover fermions for $a\sim 0.08$~fm and $m_\pi\sim
  350$~MeV (open cyan diamonds)~\cite{Brommel:2008tc}; and iv)
  $N_f{=}2{+}1$ with DWF valence on a staggered sea for $a=0.124$~fm
  and $m_\pi=356$~MeV  (open orange circles)~\cite{Bratt:2010jn}. }
\label{fig:A20A20tilde_various}
\end{figure}

Recent results on $A_{20}$ and $\tilde{A}_{20}$   from a number 
of groups using different
discretization schemes are shown  in
Fig.~\ref{fig:A20A20tilde_all}. We limit ourselves to results
extracted from fitting to the ratio given in Eq.~(\ref{plateau})
taking a source-sink separation of 1~fm to 1.2~fm. 
Once more, there is an overall agreement among these lattice data indicating that
 cut-off effects are small for lattice spacings  $\stackrel{<}{\sim} 0.1$fm,
for the  improved actions used.
The decrease seen using TMF at $m_\pi=213$~MeV is corroborated by other
recent  results at near physical pion masses: for  $\langle x\rangle_{u-d}$ 
results from Ref.~\cite{Bali:2012av}
 using clover-improved fermions at $m_\pi=157$~MeV and $Lm_\pi=2.74$,
as well as, from Ref.~\cite{Green:2012rr} using $N_f{=}2{+}1$ 
flavors of tree-level clover-improved Wilson fermions coupled to
double HEX-smeared gauge fields at $m_\pi=149$~MeV and $Lm_\pi=4.2$,
also decrease towards the physical value. Furthermore, for the latter
case, three sink-source separations up to 1.4~fm were utilized to
apply the summation method reducing the value shown in
Fig.~\ref{fig:A20A20tilde_all} further to bring it into
agreement with the experimental one~\cite{Green:2012ud}. Note that this
is opposite to what was found for $g_A$ where its value decreased further
away from the experimental value.  The agreement between the
 values found in Refs.~\cite{Bali:2012av} and \cite{Green:2012rr},
despite the different  volumes, indicates that the volume
dependence of this quantity is small, again different from what
was claimed in Ref.~\cite{Horsley:2013ayv} for $g_A$.  
In Ref.~\cite{Green:2012rr} it was demonstrated that contributions from
 excited states
increase  as the pion mass decreases towards its
physical value indicating that excited state contamination
may explain  the discrepancy between lattice results and the
experimental value. Further studies of excited state contamination at
near physical pion mass will be essential in order to establish this
conclusion. 

The $Q^2$-dependence of $A_{20}(Q^2)$ and $\tilde{A}_{20}(Q^2)$ is
shown in Fig.~\ref{fig:A20A20tilde_tmf} for our two $N_f{=}2{+}1{+}1$ ensembles and
for the $N_f{=}2$ ensemble with the smallest available mass, namely 262~MeV. 
Since strange and charm quark effects have been shown to be small, one can study
the dependence on the pion mass by comparing with results obtained using
$N_f{=}2$ TMF. 
As the pion mass decreases from 373~MeV to 262~MeV there is no significant change
in the values of  $A_{20}(Q^2)$ and $\tilde{A}_{20}(Q^2)$  over the whole
$Q^2$ range.  Reducing the pion mass further to 213~MeV leads to a larger decrease in the values of both
$A_{20}(Q^2)$ and $\tilde{A}_{20}(Q^2)$ indicating that near the physical regime
the pion mass dependence becomes stronger. Such a pion mass dependence is what
one would expect if the lattice QCD data at $Q^2=0$ are to agree with the experimental value.
In Fig.~\ref{fig:A20A20tilde_various} we compare our results using TMF to
hybrid results and,  for $A_{20}(Q^2)$, we also
include $N_f{=}2$ clover at  similar pion masses. 
There is  an overall agreement between clover and TMF for $A_{20}(Q^2)$,  whereas
the hybrid data are somewhat lower. The fact that they are renormalized  perturbatively  might explain their lower values. 

\begin{figure}
{\includegraphics[scale=0.35]{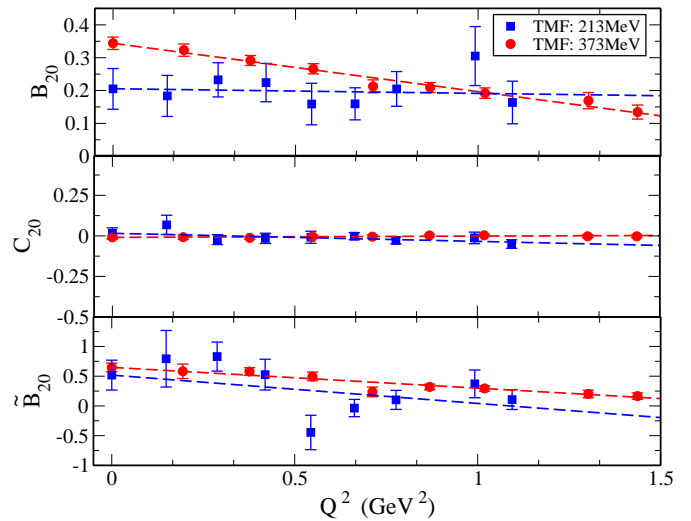}}
\caption{The $Q^2$-dependence of $B_{20}(Q^2),\,C_{20}(Q^2)$ and
$\tilde{B}_{20}(Q^2)$ for $N_f{=}2{+}1{+}1$ computed at $\beta=1.95$
($m_\pi=373$ MeV) and $\beta=2.10$ ($m_\pi=213$) MeV. The dashed lines
show the  linear fits to $B_{20}(Q^2),\,C_{20}(Q^2)$ and
$\tilde{B}_{20}(Q^2)$ to extract the value at $Q^2=0$ shown here.}
\label{fig:GFFscompare}
\end{figure}

Before closing this section we present in Fig.~\ref{fig:GFFscompare} results for 
$B_{20}(Q^2),\,C_{20}(Q^2),\,\tilde B_{20}(Q^2)$ for the two $N_f{=}2{+}1{+}1$ ensembles.
All these three GFFs can not be extracted at $Q^2{=}0$ directly from the 
matrix element and therefore we must extrapolate them using
an Ansatz to fit the $Q^2$-dependence. 
 We performed two types of fits: a  linear and a dipole fit. Note that
for small $Q^2$ the two are equivalent.
It was generally found that a linear fit describes well the data 
with smaller errors on the fit parameters.
We therefore use the fitted values extracted from the linear fit summarized in
Table~\ref{tab:B20C20B20tilde}. $C_{20}(Q^2)$ is consistent with zero for
all values of $Q^2$.  
  \begin{table}[h]
\begin{center}
\begin{tabular}{c|c|c|c}
\hline\hline
$\beta$ & $B_{20}(0)$ (GeV)& $C_{20}(0)$ & $\tilde B_{20}(0)$
\\\hline
1.95 &  0.344(19)  & -0.009(09)   & 0.648(71)     \\
2.10 &  0.205(62)  & 0.016(34)    & 0.518(251)     \\
\hline
\end{tabular}
\caption{Results on $B_{20}(Q^2=0),\,C_{20}(Q^2=0)$ and
$\tilde{B}_{20}(Q^2=0)$
by fitting to a linear $Q^2$-dependence.}
\label{tab:B20C20B20tilde}
\end{center}
\end{table}

\section{Proton Spin}
How much of the proton spin is carried by the quarks is a question that
is under study ever since the results of  the European Muon Collaboration 
(EMC) claimed that the quarks carried only a small fraction of the proton spin~\cite{Ashman:1987hv}. This  became known as the ``proton spin crisis''. 
It was proposed that  gluons in a polarized proton 
would carry a fraction of the spin, which however would be unnaturally large
if it were to resolve the EMC spin crisis. It is now understood that the resolution of this puzzle requires to take into account the non-perturbative structure of the proton~\cite{Thomas:2009gk}.  In order to use our lattice results
to  obtain information on the spin content of the nucleon we need to 
evaluate, besides the isovector moments, the isoscalar moments $A_{20}^{u+d}$ and
$B_{20}^{u+d}$ since  the total  angular momentum of a quark in the nucleon is given
by 
\be J^q=\frac{1}{2}\left ( A_{20}^q (0)+
B_{20}^q(0) \right)\,.
\ee
As already discussed, the total angular momentum $J^q$  can be further decomposed into its orbital angular momentum $L^q$
and its spin component $\Delta\Sigma^q$ as
\be
J^q=\frac{1}{2}\Delta\Sigma^q +L^q\,.
\ee
The spin carried by the u- and d- quarks is
determined using $\Delta\Sigma^{u+d}=\tilde{A}_{10}^{u+d}$, and therefore
we need the isoscalar axial charge.
The isoscalar quantities take contributions from the disconnected
diagram, which are notoriously difficult to calculate and are neglected
in most current evaluations of GFFs. These contributions are
currently being computed using improved stochastic techniques~\cite{Alexandrou:2012gz,Alexandrou:2012zz}. Under the assumption that these
are small we may extract information on the fraction of the nucleon spin carried by quarks.

\begin{figure}
{\includegraphics[scale=0.35]{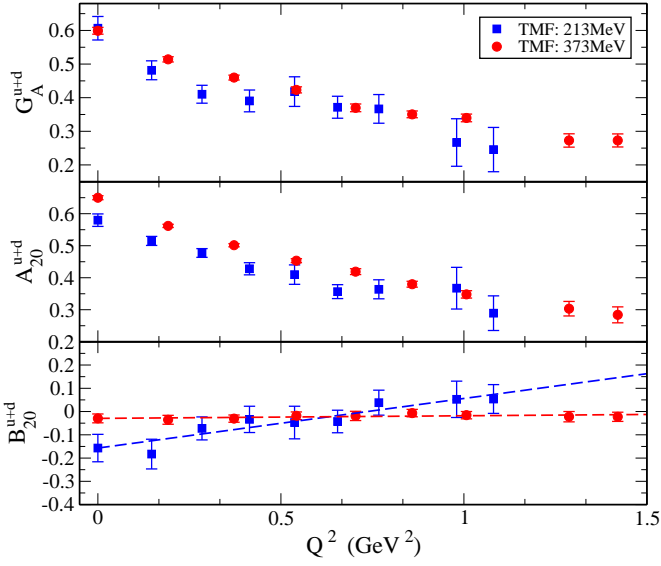}}
\caption{The $Q^2$-dependence of the isoscalar $G_A(Q^2)$,
  $A_{20}(Q^2)$ and $B_{20}(Q^2)$  for $N_f{=}2{+}1{+}1$ computed at
  $\beta=1.95$ ($m_\pi=373$ MeV) and $\beta=2.10$ ($m_\pi=213$) MeV.  }
\label{fig:GA_A20_B20_IS}
\end{figure}

In Fig.~\ref{fig:GA_A20_B20_IS} we show our results for
the isoscalar $G_A(Q^2)^{u+d}$, $A_{20}(Q^2)^{u+d}$, $B_{20}(Q^2)^{u+d}$
and $C_{20}(Q^2)^{u+d}$ for the two $N_f{=}2{+}1{+}1$ ensembles analyzed
in this work. It was shown using the $N_f{=}2$ ensembles at three
lattice spacings smaller than 0.1~fm~\cite{Alexandrou:2011nr} that
cut-off effects are small. We expect a similar behaviour for our $N_f{=}2{+}1{+}1$
ensembles.  Therefore, we
perform a chiral extrapolation using directly all our lattice data for
the $N_f{=}2$ and $N_f{=}2{+}1{+}1$ ensembles. 
Having both isoscalar and isovector quantities we can extract the
 angular momentum $J^u$ and $J^d$ carried by the u- and d- quarks. In order
to extract these quantities we need to know the value of $B_{20}$ at
$Q^2=0$. As explained already,
   one has to extrapolate the lattice results using an Ansatz for the $Q^2$-dependence to extract $B_{20}$ at $Q^2=0$ and  two  ans\"atze were considered for the $Q^2$-dependence, a dipole
 and a linear form. For
the linear fit we use two fitting ranges one up to
$Q^2=0.25$~GeV$^2$ and the other up to $Q^2=4$~GeV$^2$. Thus the 
extrapolation introduces model dependence in the extraction of the
quark spin $J^q$. The values of $B_{20}$ extracted using these three
ans\"atze are consistent, with the dipole fit resulting in parameters that carry large
errors. In extracting the  angular momentum we thus use the data extracted using the extended
range linear fit and given in Table~\ref{tab:B20C20B20tilde}.

\begin{figure}
{\includegraphics[scale=0.3,angle=-90]{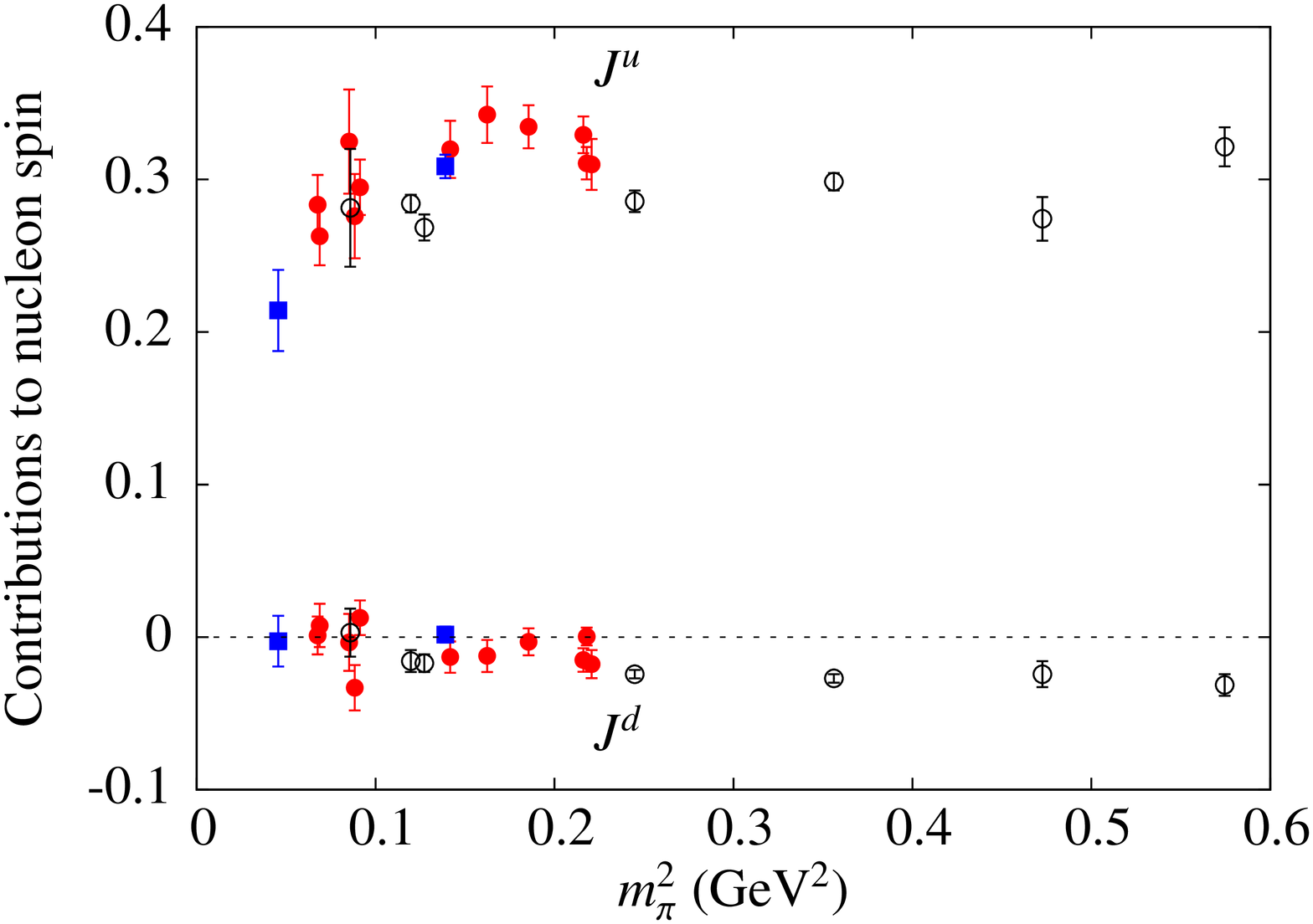}}
{\includegraphics[scale=0.3,angle=-90]{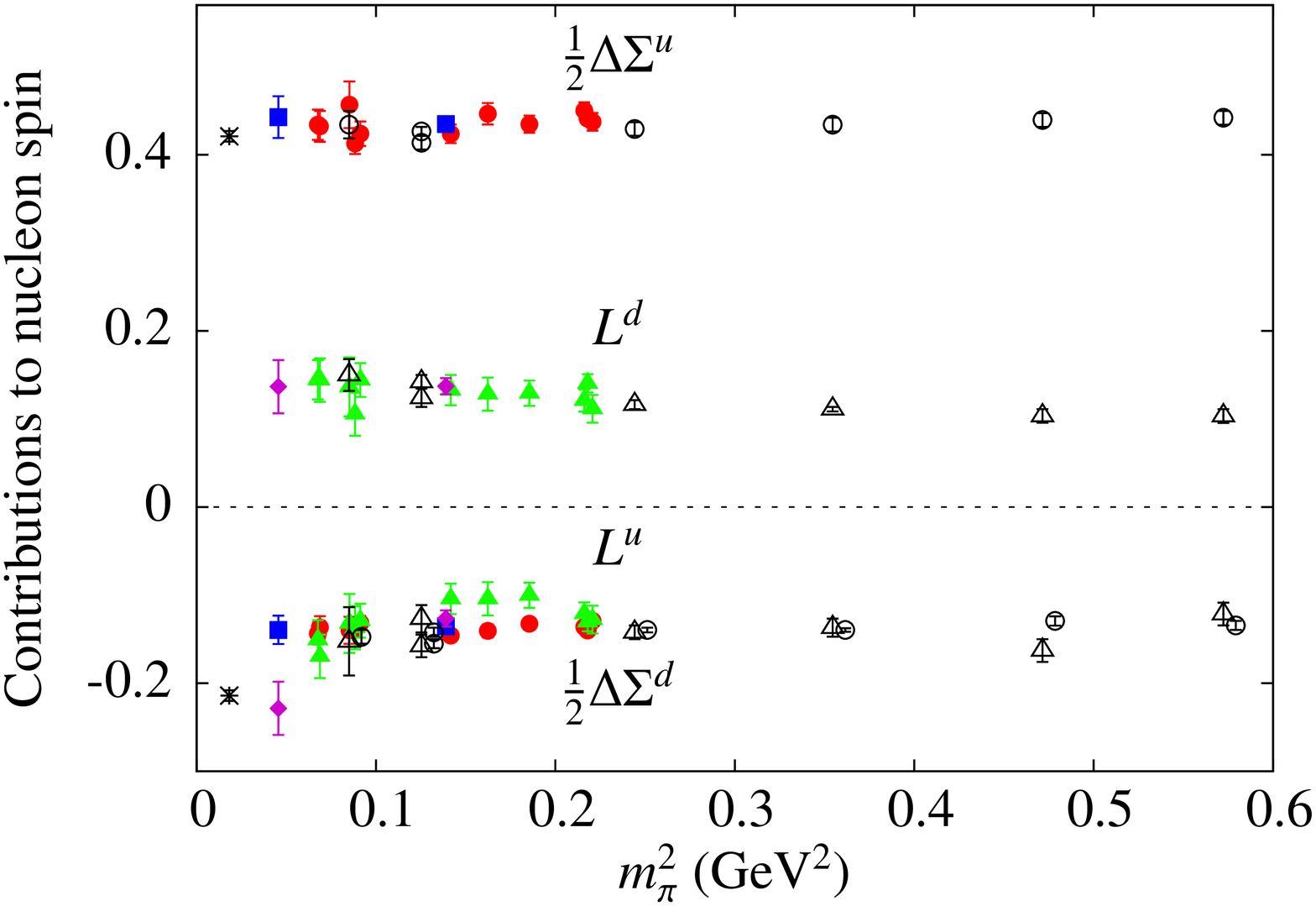}}
\caption{Comparison of TMF results (filled symbols) to
  those using a hybrid action~\cite{Bratt:2010jn} (open symbols). The
  upper panel shows the  angular momentum $J^u$  and $J^d$  for u- and
  d- quarks respectively (blue filled squares for $N_f{=}2{+}1{+}1$
  and filled red circles for $N_f{=}2$). The lower panel shows the
   quark spin (same symbols as for $J^q$) and 
 the  orbital angular momentum (filled green triangles for $N_f{=}2$ and filled
  magenta diamonds for $N_f{=}2{+}1{+}1$). 
The errors are determined by carrying out a superjacknife
analysis described in Ref.~\cite{Bratt:2010jn}. The experimental value 
of $\Delta\Sigma^{u,d}$ is shown by the asterisks and are taken 
from the HERMES 2007 analysis~\cite{Airapetian:2007mh}.}
\label{fig:J comparison}
\end{figure}

 We first compare in Fig.~\ref{fig:J comparison} our
results for the  u- and d- quark angular momentum $J^q$,  spin $\Delta \Sigma^q$
and orbital angular momentum $L^q$ to those obtained using the hybrid
action of Ref.~\cite{Bratt:2010jn}. As can be seen, the lattice data are in
agreement within our statistical errors indicating that lattice
artifacts are smaller than the current statistical errors, also for these quantities.
\begin{figure}
{\includegraphics[scale=0.3,angle=-90]{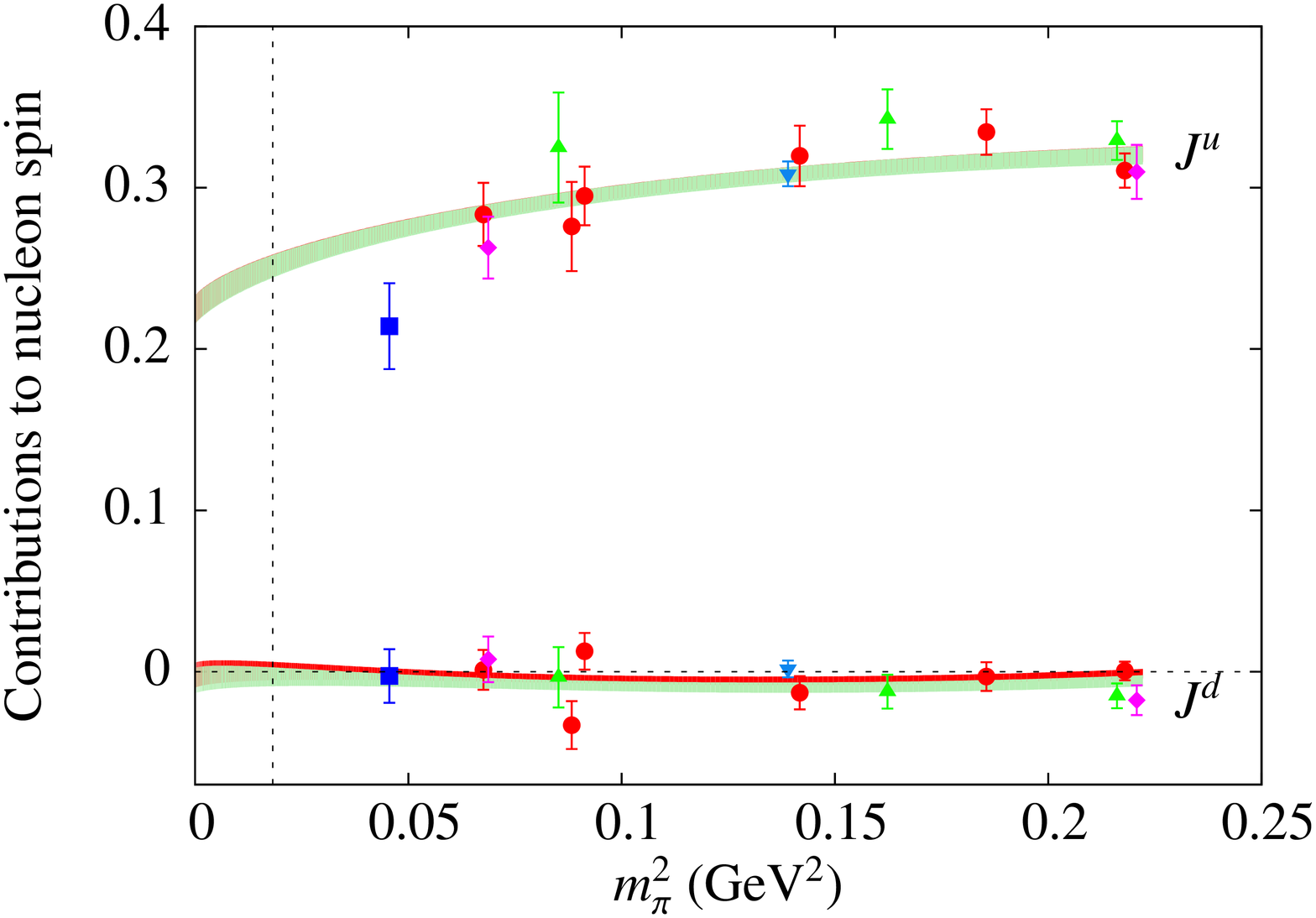}}
{\includegraphics[scale=0.3,angle=-90]{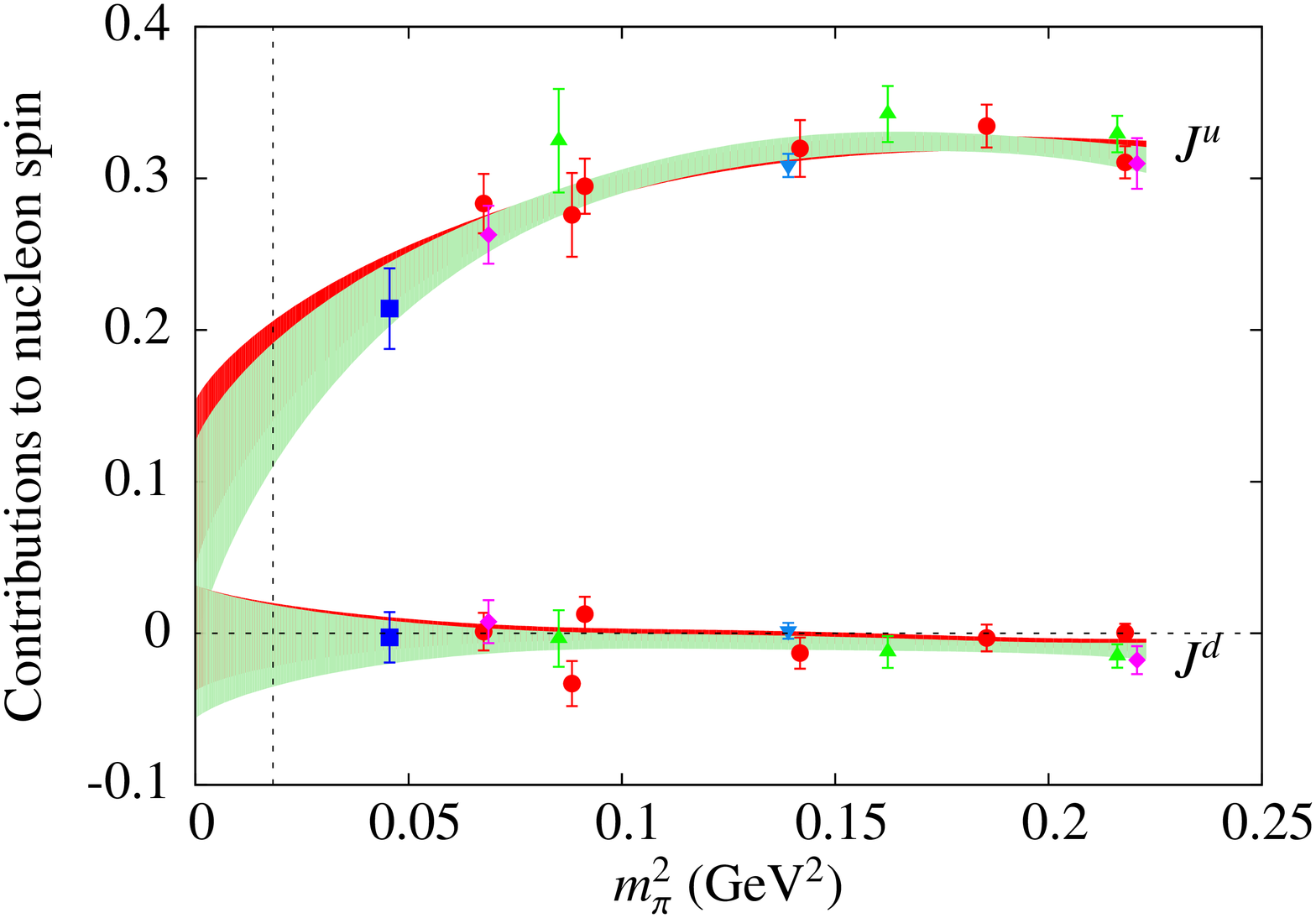}}
\caption{Chiral extrapolation using CB$\chi$PT (upper)
 and HB$\chi$PT (lower) for the  angular momentum carried by the u-and d-
 quarks. The red band is the chiral fit using the data for
 $B_{20}(Q^2=0)$ obtained by a linear extrapolation of $B_{20}(Q^2)$
 using  $Q^2$ values up to $Q^2=4$~GeV$^2$ whereas the green band is the fit
using values of $B_{20}(0)$ extracted from a linear extrapolation of
$B_{20}(Q^2)$ using $Q^2$ values up to $\sim 0.25$~GeV$^2$. The  data
shown in the plot are obtained from the extended linear $Q^2$
extrapolation. Filled red circles are data for $N_f{=}2$ at $\beta=3.9$,
filled green triangles for $N_f{=}2$ at $\beta=4.05$, filled magenta
diamonds for $N_f{=}2$ at $\beta=4.2$, filled light blue inverted triangle
for $N_f{=}2{+}1{+}1$ at $\beta=1.95$ and filled blue square for $N_f{=}2{+}1{+}1$
at $\beta=2.10$.} 
\label{fig:J CB HB}
\end{figure}
\begin{figure}
{\includegraphics[scale=0.3,angle=-90]{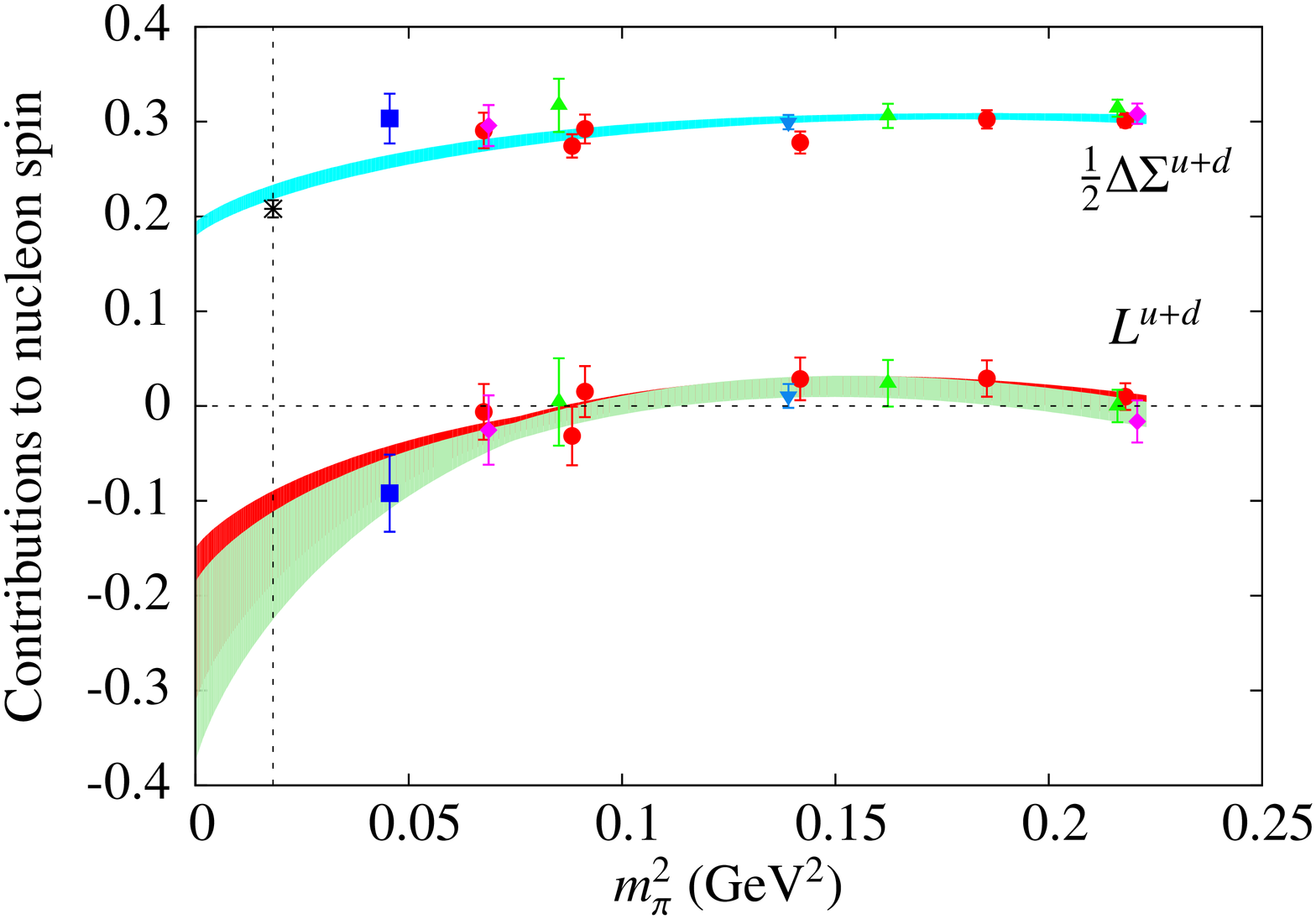}}
{\includegraphics[scale=0.3,angle=-90]{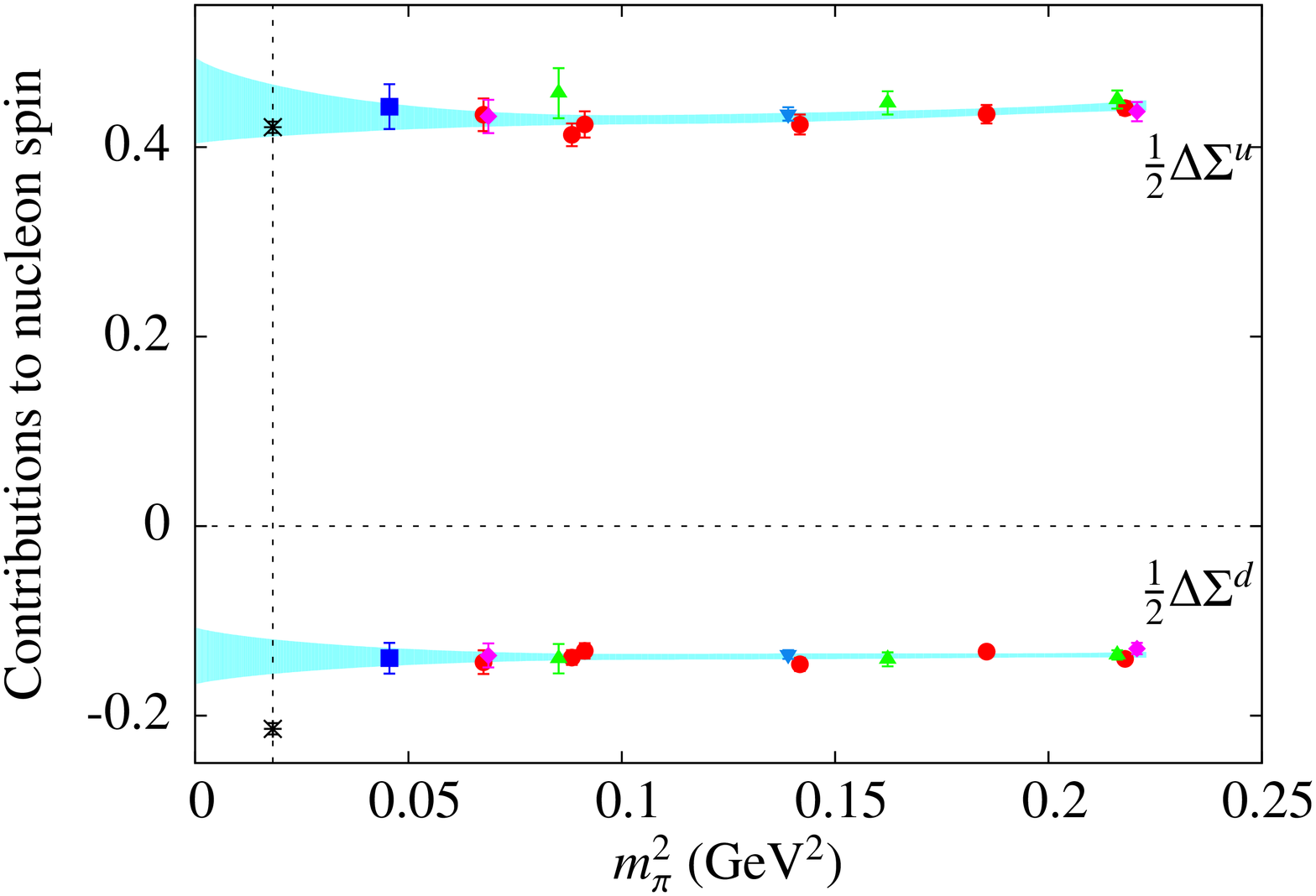}}
{\includegraphics[scale=0.3,angle=-90]{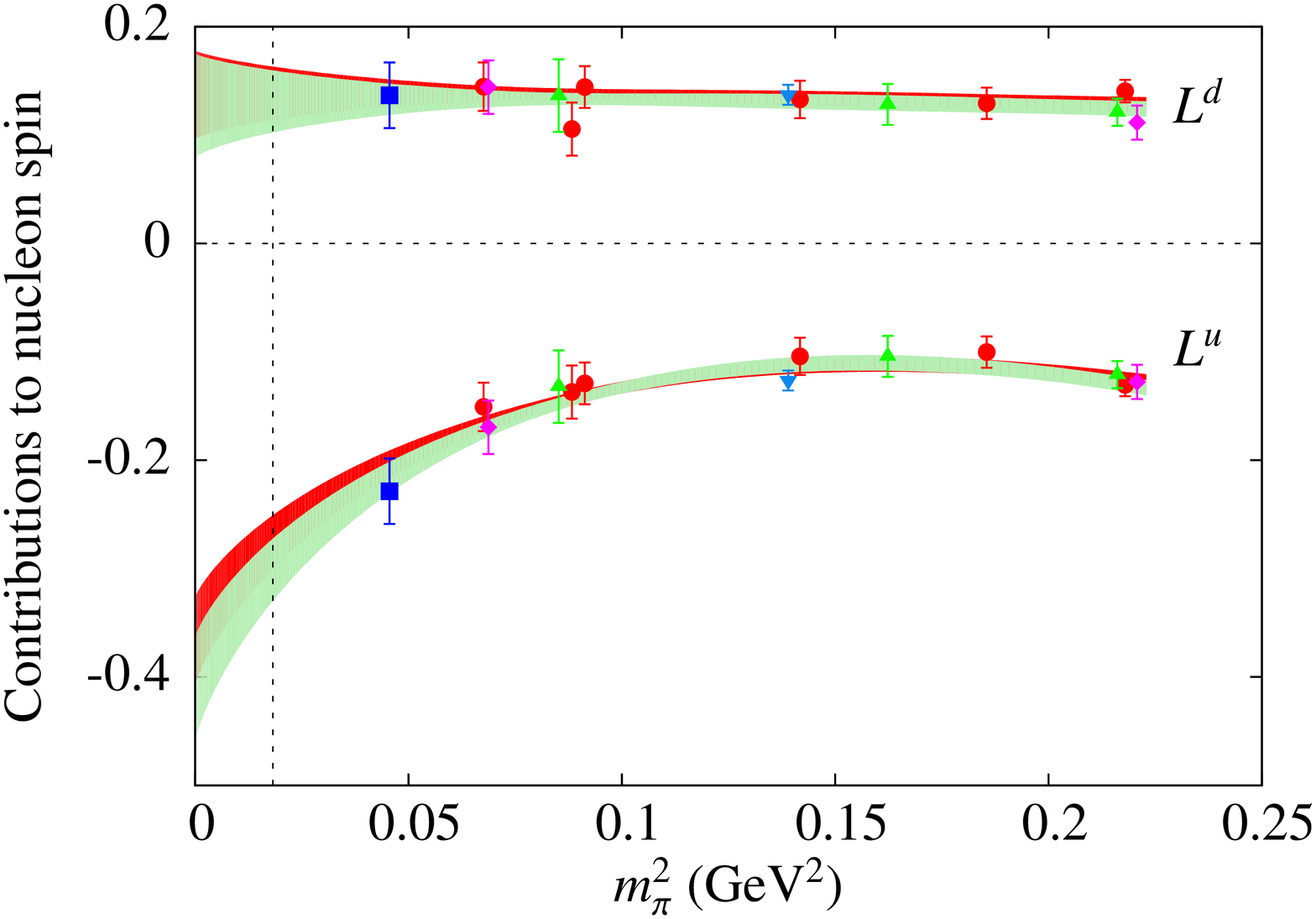}}
\caption{Chiral extrapolation using HB$\chi$PT.
 The upper graph shows the  spin and orbital angular
momentum carried by u- and d- quarks, whereas the middle and
lower graphs show the spin and orbital angular momentum carried
separately by the u- and d- quarks. The errors are determined through
a superjacknife analysis. The physical points, shown by the asterisks
are from the HERMES 2007 analysis~\cite{Airapetian:2007mh}. The
notation is the same as that in Fig.~\ref{fig:J CB HB}.}
\label{fig:L and S}
\end{figure}
In order to get an approximate value for these observables at the
physical point we perform a chiral extrapolation using heavy baryon
chiral perturbation theory (HB$\chi$PT). Combining the expressions for
$A_{20}$ and $B_{20}$~\cite{Arndt:2001ye,Detmold:2002nf} in the
isoscalar and isovector cases we obtain the following form for the
 angular momentum
\be 
J^q = a_0^q\frac{m_\pi^2}{(4\pi f_\pi)^2}  \ln\frac{m_\pi^2}{\lambda^2}  +
a_1^q m_\pi^2 + a_2^q\>,
\label{chiral J}
\ee
and take $\lambda^2=1$~GeV$^2$. We also carry out a chiral fit
using $\Op(p^2)$ covariant baryon chiral perturbation theory
(CB$\chi$PT)~\cite{Dorati:2007bk}. All the expressions are collected
in Appendix B for completeness. As noted  these chiral extrapolations
are to give an  indicative idea of what one might obtain since
their range of validity may require using pion masses closer to the
physical point.

In order to correctly estimate the errors both on the data points
and on the error bands, we  apply an extended 
 version of the standard jackknife error procedure known as superjackknife analysis~\cite{Bratt:2010jn}. This
generalized method is applicable for analyzing data computed on several 
gauge ensembles. This is needed for carrying out the chiral extrapolations
for the  angular momentum
 $J^q$, orbital  angular momentum $L^q$ and  spin $\Delta\Sigma^q$. Although, 
there is no
correlation among data sets from different  gauge 
 ensembles, the  data within each
ensemble  are correlated. This analysis method 
 allows us to consider a different number of lattice QCD measurements for each
ensemble   taking into account correlations within each ensemble correctly.
 It should be apparent that the superjackknife
reduces to the standard jackknife analysis in the case of a single ensemble.

In Fig.~\ref{fig:J CB HB} we show the chiral fits for $J^q$. In the
upper panel we show the chiral extrapolation using CB$\chi$PT and in
the lower the extrapolation using HB$\chi$PT. Both have the same
qualitative behavior yielding a much smaller contribution to the  angular momentum
from the d-quark than that from the u-quarks. In the plot we also
show the band of allowed values if the fit were performed on data that used
the $Q^2=0$ extrapolated values of $B_{20}$ from the limited range linear fit.
As can be seen, the two bands are consistent. Had we used a dipole Ansatz for the
$Q^2=0$ extrapolation, the error band would also be consistent but 
 much 
larger, especially for smaller pion masses, where there are no lattice  data. Therefore, for the rest of the discussion we only show the extrapolation bands obtained using the limited and full $Q^2$ range linear fits.  
These results are in qualitative agreement with the chiral extrapolations
 using the data obtained with the hybrid action~\cite{Bratt:2010jn}.

In Fig.~\ref{fig:L and S} we show separately the orbital angular
momentum and spin carried by the u- and d- quarks. The 
total orbital  angular momentum carried by the quarks tends to small negative values as 
we approach the physical point. This is a crucial result and it would be
important to perform a calculation at lower pion mass to confirm that this
trend towards negative values remains~\footnote{There was a mistake in
  the extraction of $L^q$ of Ref.~\cite{Alexandrou:2011nr} using
  $N_f{=}2$ twisted mass fermions, which is corrected here.}. After
chiral extrapolation, the value obtained at the physical point is
consistent with zero in agreement with the result by LHPC. We summarize
the values for the  angular momentum, orbital angular momentum and spin in
the proton at 
the smallest pion mass, namely at $m_\pi=213$~MeV  in Table~\ref{tab:spin parameters}.
The pion mas dependence of   $\Delta \Sigma^u$ and $\Delta \Sigma^d$  is weak as can be seen in Fig.~\ref{fig:J comparison} and if one
assumes that this continuues up to the physical pion mass then  $\Delta \Sigma^u$ agrees with the experimental value
whereas   $\Delta \Sigma^d$ is less negative.
As already pointed out, results closer to the physical pion mass will be essential to resolve such discrepancies. In addition, the computation of the  disconnected diagrams will eliminate a remaining  systematic error and will enable
us to have final results on the spin carried by the quarks and consequently on the gluon contribution to the nucleon spin.

\begin{table}
\begin{center}
\begin{tabular}{|c|c|c|}
\hline
\hline
        		&   $m_\pi=213$ MeV   & experiment \\
\hline                
 $J^{u-d}$		&   0.217(32)         &              \\
 $J^{u+d}$      	&   0.211(30)         &      \\
 $J^u    $ 	     	&   0.214(27)	      &          \\
 $J^d    $      	&   -0.003(17)        &         \\
 $\Delta\Sigma^{u-d}/2$	&   0.582(31)	      & 0.634(2) \\
 $\Delta\Sigma^{u+d}/2$  &   0.303(26)        & 0.208(9)  \\
 $\Delta\Sigma^u/2$     &   0.443(24)	      & 0.421(6)  \\
 $\Delta\Sigma^d/2$     &   -0.140(16)        &-0.214(6)  \\
 $L^{u-d}$               &   -0.365(45)        &         \\
 $L^{u+d}$             	&   -0.092(41)         &       \\
 $L^u$                	&  -0.229(30)         &         \\
 $L^d$               	&    0.137(30)       &          \\
\hline\hline
\end{tabular}
\caption{Values of nucleon spin observables at $m_\pi=213$~MeV, the smallest pion mass available in our
LQCD simulations,  and from experiment~\cite{Airapetian:2007mh}. The error on the LQCD values are only statistical.} 
\label{tab:spin parameters}
\end{center}
\end{table}


\section{Conclusions}
We have performed an analysis on the generalized form factors $G_E(Q^2)$,
$G_M(Q^2)$, $G_A(Q^2)$, $G_p(Q^2)$,
$A_{20}(Q^2)$, $B_{20}(Q^2)$, $C_{20}(Q^2)$, $\tilde A_{20}(Q^2)$ and
$\tilde B_{20}(Q^2)$, extracted from the nucleon matrix elements of
the local and one-derivative vector and axial-vector operators using
$N_f{=}2{+}1{+}1$ flavors of twisted mass fermions.
Our results are non-perturbatively renormalized and they are presented
in the $\overline{\rm MS}$ scheme at a scale of 2 GeV.
The comparison of the results using $N_f{=}2$ and $N_f{=}2{+}1{+}1$
twisted mass fermions with the results obtained using other
discretizations show an overall agreement for pion masses down to
about 200~MeV. The compatibility of $N_f{=}2$ data
with those including a dynamical strange and a charm quark is an
indication that any systematic effect of strange and charm sea quark effects on these quantities for which  disconnected contributions were neglected is small.
The twisted mass fermion results on the axial nucleon charge remain
smaller than the experimental value. The recent results using
$N_f{=}2$~\cite{Horsley:2013ayv} and $N_f{=}2{+}1$~\cite{Green:2012ud}
clover-improved fermions near the physical pion mass are somewhat in
conflict with each other and hard to interpret in a consistent
way. Therefore, further investigation is required to resolve the
issue. For the unpolarized isovector momentum fraction lattice results
show a decrease as we approach the physical pion mass with indications
of excited state contamination that needs further investigation.

We also analyze the corresponding isoscalar quantities using directly
our lattice data. Of particular interest here is to extract results
that shed light on the spin content of the nucleon. Assuming that the
disconnected contributions to the isoscalar quantities are small we
can extract the spin carried by the quarks in the nucleon. For the
chiral extrapolations of these quantities we use HB$\chi$PT and
CB$\chi$PT theory applied to all our $N_f{=}2$ and $N_f{=}2{+}1{+}1$
data. We find that the spin carried by the d-quark is almost zero
whereas the u-quarks carry about 50\% of the nucleon spin. This
result is consistent with other lattice calculations~\cite{Bratt:2010jn}.

\section*{Acknowledgments}
We would like to thank all members of ETMC for a
very constructive and enjoyable collaboration and for the many fruitful
discussions that took place during the development of this work.

Numerical calculations have used HPC resources from John von
Neumann-Institute for Computing on the Jugene systems at the research
center in J\"ulich through the PRACE allocation, 3rd regular call.
It also used the Cy-Tera facility of the Cyprus Institute under the  project Cy-Tera (NEA Y$\Pi$O$\Delta$OMH/$\Sigma$TPATH/0308/31), first access call  (project lspro113s1).
We thank the staff members for their kind and
sustained support.
This work is supported in part by the Cyprus Research Promotion
Foundation under contracts KY-$\Gamma$/0310/02/ and
TECHNOLOGY/$\Theta$E$\Pi$I$\Sigma$/0311(BE)/16, and the Research Executive Agency of the European Union under Grant Agreement number PITN-GA-2009-238353 (ITN STRONGnet). This work is also supported in part by the DFG Sonderforschungsbereich/Transregio SFB/TR9.
K. J. was supported in part by the Cyprus Research Promotion
Foundation under contract $\Pi$PO$\Sigma$E$\Lambda$KY$\Sigma$H/EM$\Pi$EIPO$\Sigma$/0311/16.

\bibliography{GPDs_ref}

\newpage

\begin{widetext}

{\bf{Appendix A: Numerical results for the isovector sector}}
In Tables~\ref{tab:results_ffs} and \ref{tab:results_Gffs} we tabulate
our results for the isovector quantities which was presented in the
main part of the paper, that is,
$G_E,\,G_M,\,G_A,\,G_p,\,A_{20},\,B_{20},\,\tilde{A}_{20}$ and
$\tilde{B}_{20}$. For completeness, we include the isoscalar
quantities $G_A^{IS},\,A_{20}^{IS}$ and $B_{20}^{IS}$, which are
required in the extraction of the orbital angular momentum and spin
component of the nucleon.
\small
\begin{table}[!h]
\begin{center}
\begin{tabular}{|c|c|c|c|c|c|c|}
\hline\hline
 $m_\pi$ (GeV) & $(Q)^2$ & $G_E$ & $G_M$ & $ G_A$ & $G_p$ & $G_A^{IS}$ \\
 (no. confs)  &  &  &   &  &  &  \\\hline
\multicolumn{4}{c}{$\beta=1.95$, $32^3\times 64$ }\\\hline
            &0.0       & 1.000(1)   & 3.930(117) & 1.141(18)   & 18.211(9.209) &0.599(15)       \\
            &0.192     & 0.734(6)   & 2.979(61)  & 0.995(14)   & 9.462(399)    &0.514(12)       \\
   0.373    &0.372     & 0.570(7)   & 2.355(46)  & 0.872(12)   & 6.116(226)    &0.460(11)       \\
  (950)     &0.542(1)  & 0.469(10)  & 1.937(47)  & 0.775(14)   & 4.512(209)    &0.423(14)       \\
            &0.704(1)  & 0.392(12)  & 1.676(57)  & 0.714(21)   & 3.117(208)    &0.370(17)       \\
            &0.859(2)  & 0.331(11)  & 1.405(46)  & 0.642(18)   & 2.591(134)    &0.350(15)       \\
            &1.007(2)  & 0.288(13)  & 1.250(53)  & 0.589(21)   & 2.134(129)    &0.340(17)       \\
            &1.287(3)  & 0.208(20)  & 0.950(79)  & 0.480(39)   & 1.441(182)    &0.273(30)       \\
            &1.420(4)  & 0.185(20)  & 0.865(85)  & 0.450(41)   & 1.249(163)    &0.273(29)       \\
 \hline
\multicolumn{4}{c}{$\beta=2.10$, $48^3\times 96$ }\\\hline
            &0.0       & 1.006(6)   & 3.855(342) & 1.164(62)   & 14.880(11.790) &0.607(52)      \\
            &0.147     & 0.722(21)  & 2.849(198) & 1.034(47)   &10.454(1.445)   &0.481(42)      \\
   0.213    &0.284     & 0.565(23)  & 2.347(142) & 0.909(42)   & 6.317(783)     &0.410(40)      \\
  (900)     &0.414(1)  & 0.430(30)  & 1.950(153) & 0.850(52)   & 5.227(699)     &0.390(49)      \\
            &0.537(1)  & 0.444(41)  & 1.622(170) & 0.690(68)   & 2.466(723)     &0.418(66)      \\
            &0.655(2)  & 0.318(29)  & 1.338(120) & 0.689(53)   & 2.628(395)     &0.371(49)      \\
            &0.768(3)  & 0.266(32)  & 1.291(136) & 0.707(71)   & 2.763(481)     &0.367(64)      \\
            &0.980(4)  & 0.218(52)  & 1.104(237) & 0.558(129)  & 2.466(701)     &0.267(106)     \\
            &1.081(5)  & 0.186(44)  & 0.686(164) & 0.437(110)  & 1.714(541)     &0.246(99)      \\
\hline
\end{tabular}
\caption{Results on the isovector $G_E,\,G_M,\,G_A$ and $G_p$ and isoscalar $G_A^{IS}$ form factors at $\beta=1.95$ ($32^3\times64$) and $\beta=2.10$ ($48^3\times96$). $G_A^{IS}(0)$ is needed to extract the spin carried by quarks in the nucleon.}
\label{tab:results_ffs}
\end{center}
\end{table}

\begin{table}
\begin{center}
\begin{tabular}{|c|c|c|c|c|c|c|c|}
\hline\hline
 $m_\pi$ (GeV) & $(Q)^2$  & $A_{20}$ & $B_{20}$ & $\tilde A_{20}$ & $\tilde B_{20}$ & $A_{20}^{IS}$ & $B_{20}^{IS}$    \\
 (no. confs)   &   &  &   &  &  & &  \\\hline
\multicolumn{4}{c}{$\beta=1.95$, $32^3\times 64$ }\\\hline
            &0.0       & 0.270(5)  & 0.344(19) & 0.302(5)   & 0.648(71)    &0.650(6)    &-0.029(19)    \\
            &0.192     & 0.242(4)  & 0.292(15) & 0.281(5)   & 0.582(121)   &0.562(5)    &-0.035(19)    \\
   0.373    &0.372     & 0.222(4)  & 0.266(15) & 0.264(5)   & 0.578(65)    &0.502(5)    &-0.030(16)    \\
  (950)     &0.542(1)  & 0.207(5)  & 0.266(15) & 0.249(6)   & 0.495(73)    &0.453(6)    &-0.018(17)    \\
            &0.704(1)  & 0.195(6)  & 0.213(20) & 0.231(7)   & 0.236(81)    &0.419(9)    &-0.019(20)    \\
            &0.859(2)  & 0.177(6)  & 0.209(15) & 0.219(7)   & 0.319(46)    &0.380(11)   &-0.007(16)    \\
            &1.007(2)  & 0.163(8)  & 0.192(16) & 0.202(9)   & 0.294(47)    &0.348(11)   &-0.016(16)    \\
            &1.287(3)  & 0.152(14) & 0.169(25) & 0.170(15)  & 0.200(63)    &0.303(23)   &-0.022(22)    \\
            &1.420(4)  & 0.134(14) & 0.134(21) & 0.164(16)  & 0.165(53)    &0.284(25)   &-0.023(20)    \\
 \hline
\multicolumn{4}{c}{$\beta=2.10$, $48^3\times 96$ }\\\hline
            &0.0       & 0.228(18)  & 0.205(62) & 0.251(19)   & 0.518(251)    &0.580(19)   &-0.157(59)    \\
            &0.147     & 0.206(14)  & 0.184(63) & 0.242(14)   & 0.793(475)    &0.515(14)   &-0.183(64)    \\
  0.213     &0.284     & 0.190(12)  & 0.233(52) & 0.247(14)   & 0.830(244)    &0.477(14)   &-0.072(49)    \\
  (900)     &0.414(1)  & 0.165(16)  & 0.224(58) & 0.229(18)   & 0.526(259)    &0.428(19)   &-0.034(57)    \\
            &0.537(1)  & 0.176(23)  & 0.159(63) & 0.204(25)   &-0.446(289)    &0.410(30)   &-0.047(70)    \\
            &0.655(2)  & 0.152(16)  & 0.159(49) & 0.180(19)   &-0.036(145)    &0.357(22)   &-0.043(48)    \\
            &0.768(3)  & 0.167(20)  & 0.205(53) & 0.181(24)   & 0.101(160)    &0.364(30)   & 0.038(54)    \\
            &0.980(4)  & 0.173(40)  & 0.305(90) & 0.164(43)   & 0.371(233)    &0.367(65)   & 0.052(78)    \\
            &1.081(5)  & 0.142(33)  & 0.164(65) & 0.122(35)   & 0.106(166)    &0.289(54)   & 0.054(62)    \\
\hline
\end{tabular}
\caption{Results on the isovector $A_{20},\,B_{20},\,\tilde A_{20}$ and $\tilde B_{20}$ and isoscalar $A_{20}^{IS}$ and $B_{20}^{IS}$ generalized  form factors at $\beta=1.95$ ($32^3\times64$) and $\beta=2.10$ ($48^3\times96$).
}
\label{tab:results_Gffs}
\end{center}
\end{table}

\normalsize

\newpage

{\bf{Appendix B: Expressions for the chiral extrapolation of the quark spin and
angular momentum}}\\[5ex]
In this Appendix we collect the expression used to extrapolate our lattice data
for the quark spin to the physical point.
Throughout, we use $\lambda^2 =$ 1~GeV$^2$, $f_\pi =0.0924$~GeV and $g_A=1.267$.

In HB$\chi$PT the expressions for $A_{20}(0)$ and $B_{20}(0)$ for the isovector combination 
are given by

\begin{eqnarray}
	A_{20}^{I=1}(0)&=&A_{20}^{I=1(0)}\left\lbrace 1 - \frac{m_\pi^2}{(4\pi f_\pi)^2}\left[(3g_A^2+1)\ln\frac{m_\pi^2}{\lambda^2} + 2g_A^2 \right]\right\rbrace + A_{20}^{I=1(2,m)} m_\pi^2 \\
	B_{20}^{I=1}(0)&=&B_{20}^{I=1(0)}\left\lbrace 1 - \frac{m_\pi^2}{(4\pi f_\pi)^2}\left[(2g_A^2+1)\ln\frac{m_\pi^2}{\lambda^2} + 2g_A^2 \right]\right\rbrace 
	 + A_{20}^{I=1(0)} \frac{m_\pi^2g_A^2}{(4\pi f_\pi)^2}\ln\frac{m_\pi^2}{\lambda^2} + B_{20}^{I=1(2,m)} m_\pi^2
\end{eqnarray}
and for the isoscalar by

\begin{eqnarray}
	A_{20}^{I=0}(0)&=&A_{20}^{I=0(0)} + A_{20}^{I=0(2,m)} m_\pi^2 \\
	B_{20}^{I=0}(0)&=&B_{20}^{I=0(0)}\left[ 1 - \frac{3g_A^2m_\pi^2}{(4\pi f_\pi)^2}\ln\frac{m_\pi^2}{\lambda^2} \right] - A_{20}^{I=0(0)}\frac{3g_A^2m_\pi^2}{(4\pi f_\pi)^2}\ln\frac{m_\pi^2}{\lambda^2} 
	+ B_{20}^{I=0(2,m)} m_\pi^2 + B_{20}^{I=0(2,\pi)}\,.
\end{eqnarray}

The spin carried by the quarks is given by the axial coupling $g_A$ or $\tilde{A}_{10}(0)$ as 
\begin{eqnarray}
	\Delta\Sigma^{u+d}&=&\tilde{A}_{10}^{u+d}=\tilde{A}_{10}^{I=0}(0) \\
	\Delta\Sigma^{u-d}&=&\tilde{A}_{10}^{u-d}=\tilde{A}_{10}^{I=1}(0)\,.
\end{eqnarray} 
The corresponding expressions for $\tilde{A}_{10}(0)$ in the isoscalar and
isovector cases are

\begin{eqnarray}
	\tilde{A}_{10}^{I=1}(0)&=&\tilde{A}_{10}^{I=1(0)}\left\lbrace 1 - \frac{m_\pi^2}{(4\pi f_\pi)^2}\left[(2g_A^2+1)\ln\frac{m_\pi^2}{\lambda^2} + g_A^2 \right]\right\rbrace + \tilde{A}_{10}^{I=1(2,m)} m_\pi^2 \\
\label{a10isov}
	\tilde{A}_{10}^{I=0}(0)&=&\tilde{A}_{10}^{I=0(0)}\left\lbrace 1 - \frac{3g_A^2m_\pi^2}{(4\pi f_\pi)^2}\left[\ln\frac{m_\pi^2}{\lambda^2} + 1 \right]\right\rbrace + \tilde{A}_{10}^{I=0(2,m)} m_\pi^2 \,.
\end{eqnarray}

For the total spin $J$ we have

\begin{eqnarray}
J^{I=0} &=& a_0^{IS}\left[ 1-\frac{3g_A^2 m_\pi^2 }{(4\pi f_\pi)^2}\ln\frac{m_\pi^2}{\lambda^2} \right] + a_1^{IS} m_\pi^2 + a_2^{IS} \\
J^{I=1} &=& a_0^{IV} \left[ 1-\frac{m_\pi^2}{(4\pi f_\pi)^2}\left( (2g_A^2+1) \ln\frac{m_\pi^2}{\lambda^2} + 2g_A^2 \right) \right] + a_1^{IV}m_\pi^2
\end{eqnarray}

and the expression for
$\Delta\Sigma^q$, $L^q$ and $J^q$ are of the form 

\begin{equation}
Q^{u,d} = a_2^{u,d} + a_1^{u,d} m_\pi^2 + a_0^{u,d} \frac{m_\pi^2}{(4\pi f_\pi)^2} \ln\frac{m_\pi^2}{\lambda^2}
\end{equation}
where $Q=J , \Delta\Sigma , L$.

We also use covariant baryon chiral perturbation theory (CB$\chi$PT) for 
$A_{20}(0)$, $B_{20}(0)$, $C_{20}(0)$ in the isovector case

\begin{eqnarray}
	A_{20}^{I=1}(0) &= &a_{20}^v + \frac{a_{20}^vm_\pi^2}{(4\pi f_\pi)^2} \left[ -(3g_A^2+1)\ln\frac{m_\pi^2}{\lambda^2} - 2g_A^2+g_A^2\frac{m_\pi^2}{m_N^{0\>2}} \left(1+3\ln\frac{m_\pi^2}{m_N^{0\>2}}\right) \right.  
\nonumber \\
	&-&\left. \frac{1}{2}g_A^2\frac{m_\pi^4}{m_N^{0\>4}}\ln\frac{m_\pi^2}{m_N^{0\>2}}+g_A^2\frac{m_\pi}{\sqrt{4m_N^{0\>2}-m_\pi^2}}\left(14-8\frac{m_\pi^2}{m_N^{0\>2}}+\frac{m_\pi^4}{m_N^{0\>4}}\right) arccos\left(\frac{m_\pi}{2m_N^0} \right)   \right] 
\nonumber \\
&+& \frac{\Delta a_{20}^v(0)g_Am_\pi^2}{3(4\pi f_\pi)^2} \left[ 2\frac{m_\pi^2}{m_N^{0\>2}} \left(1+3\ln\frac{m_\pi^2}{m_N^{0\>2}} \right) - \frac{m_\pi^4}{m_N^{0\>4}}\ln\frac{m_\pi^2}{m_N^{0\>2}} + \frac{2m_\pi(4m_N^{0\>2}-m_\pi^2)^{\frac{3}{2}}}{m_N^{0\>4}} \right. \nonumber\\ 
&\times&\left.  {\rm arccos}\left(\frac{m_\pi}{2m_N^0}\right) \right] + 4m_\pi^2\frac{c_8^{(\lambda)}}{m_N^{0\>2}} + \mathcal{O}(p^3) \\
B_{20}^{I=1}(0) &=& b_{20}^v\frac{m_N(m_\pi)}{m_N^0} + \frac{a_{20}^vg_A^2m_\pi^2}{(4\pi f_\pi)^2}\left[\left(3+\ln\frac{m_\pi^2}{m_N^{0\>2}}\right)-\frac{m_\pi^2}{m_N^{0\>2}}\left(2+3\ln\frac{m_\pi^2}{m_N^{0\>2}}\right)   \right. 	\nonumber \\
	& +& \left.\frac{m_\pi^4}{m_N^{0\>4}}\ln\frac{m_\pi^2}{m_N^{0\>2}} -\frac{2m_\pi}{\sqrt{4m_N^{0\>2}-m_\pi^2}}\left(5-5\frac{m_\pi^2}{m_N^{0\>2}}+\frac{m_\pi^4}{m_N^{0\>4}}\right)       {\rm arccos}\left(\frac{m_\pi}{2m_N^0}\right) \right]+ \mathcal{O}(p^3) \\
C_{20}^{I=1}(0) &=& c_{20}^v\frac{m_N(m_\pi)}{m_N^0} + \frac{a_{20}^vg_A^2m_\pi^2}{12(4\pi f_\pi)^2}\left[ -1 + 2\frac{m_\pi^2}{m_N^{0\>2}}\left(1+\ln\frac{m_\pi^2}{m_N^{0\>2}}\right) \right. 	\nonumber \\
	& -& \left.\frac{m_\pi^4}{m_N^{0\>4}}\ln\frac{m_\pi^2}{m_N^{0\>2}} +\frac{2m_\pi}{\sqrt{4m_N^{0\>2}-m_\pi^2}}\left(2-4\frac{m_\pi^2}{m_N^{0\>2}}+\frac{m_\pi^4}{m_N^{0\>4}}\right)       {\rm arccos}\left(\frac{m_\pi}{2m_N^0}\right) \right]+ \mathcal{O}(p^3)
\end{eqnarray}


and  the isoscalar case

\begin{eqnarray}
	A_{20}^{I=0}(0) &=& a_{20}^s + 4m_\pi^2\frac{c_9}{m_N^{0\>2}} - \frac{3a_{20}^s g_A^2 m_\pi^2}{(4\pi f_\pi)^2} \left[\frac{m_\pi^2}{m_N^{0\>2}} +  \frac{m_\pi^2}{m_N^{0\>2}}\left(2-\frac{m_\pi^2}{m_N^{0\>2}}\right)\ln\frac{m_\pi}{m_N^0}  \right.  
\nonumber \\
	&+&\left.\frac{m_\pi}{\sqrt{4m_N^{0\>2}-m_\pi^2}}\left(2-4\frac{m_\pi^2}{m_N^{0\>2}}+\frac{m_\pi^4}{m_N^{0\>4}}\right) {\rm  arccos}\left(\frac{m_\pi}{2m_N^0} \right)   \right] + \mathcal{O}(p^3) \\
B_{20}^{I=0}(0) &=& b_{20}^s\frac{m_N(m_\pi)}{m_N^0} - \frac{3a_{20}^sg_A^2m_\pi^2}{(4\pi f_\pi)^2}\left[\left(3+\ln\frac{m_\pi^2}{m_N^{0\>2}}\right)-\frac{m_\pi^2}{m_N^{0\>2}}\left(2+3\ln\frac{m_\pi^2}{m_N^{0\>2}}\right)    \right.	
\nonumber \\
&+&	\left. \frac{m_\pi^4}{m_N^{0\>4}}\ln\frac{m_\pi^2}{m_N^{0\>2}} -\frac{2m_\pi}{\sqrt{4m_N^{0\>2}-m_\pi^2}}\left(5-5\frac{m_\pi^2}{m_N^{0\>2}}+\frac{m_\pi^4}{m_N^{0\>4}}\right)       {\rm arccos}\left(\frac{m_\pi}{2m_N^0}\right) \right] \\
C_{20}^{I=0}(0) &= &c_{20}^s\frac{m_N(m_\pi)}{m_N^0} - \frac{a_{20}^s g_A^2m_\pi^2}{4(4\pi f_\pi)^2}\left[ -1 + 2\frac{m_\pi^2}{m_N^{0\>2}}\left(1+\ln\frac{m_\pi^2}{m_N^{0\>2}}\right) \right. 	\nonumber \\
	&-&\left. \frac{m_\pi^4}{m_N^{0\>4}}\ln\frac{m_\pi^2}{m_N^{0\>2}} +\frac{2m_\pi}{\sqrt{4m_N^{0\>2}-m_\pi^2}}\left(2-4\frac{m_\pi^2}{m_N^{0\>2}}+\frac{m_\pi^4}{m_N^{0\>4}}\right)       {\rm arccos}\left(\frac{m_\pi}{2m_N^0}\right) \right]\,.
\end{eqnarray}

We then extract the total spins using
\beq
J^{u+d}&=&\frac{1}{2}\left(A_{20}^{I=0}(0)+B_{20}^{I=0}(0) \right) \nonumber \\
J^{u-d}&=&\frac{1}{2}\left(A_{20}^{I=1}(0)+B_{20}^{I=1}(0)\right)\,.
\eeq
\end{widetext}
\end{document}